\documentclass[epj]{svjour}
\usepackage{epsfig,array}
\usepackage{multirow,color}
\usepackage{epsf,graphicx}
\usepackage{times,latexsym,euscript,amstext,amssymb,amsbsy,amsmath,amsfonts}
\usepackage{mdwlist}
\usepackage{slashbox,multirow}

\newcommand{\be}{\begin{eqnarray}}
\newcommand{\ee}{\end{eqnarray}}
\newcommand{\bc}{\begin{center}}
\newcommand{\ec}{\end{center}}

\begin{document}

\title{Nucleon resonances in the fourth resonance region}
\titlerunning{Nucleon resonances in the fourth resonance region}
\author{
A.V.~Anisovich\inst{1,2} \and
E.~Klempt\inst{1}        \and
V.A.~Nikonov\inst{1,2}   \and
A.V.~Sarantsev\inst{1,2} \and
U.~Thoma\inst{1}
}

\authorrunning{A.V.~Anisovich \it et al.}

\institute{Helmholtz-Institut f\"ur Strahlen- und Kernphysik,
Universit\"at Bonn, Germany \and
Petersburg Nuclear Physics Institute, Gatchina, Russia}

\date{Received: \today / Revised version:}

\abstract{Nucleon and $\Delta$ resonances in the fourth resonance
region are studied in a multichannel partial-wave analysis which
includes nearly all available data on pion- and photo-induced
reactions off protons. In the high-mass range, above 1850\,MeV,
several alternative solutions yield a good description of the data.
For these solutions, masses, widths, pole residues and
photo-couplings are given. In particular, we find evidence for
nucleon resonances with spin-parities $J^P=1/2^+\cdots 7/2^+$.  For
one set of solutions, there are four resonances forming naturally a
spin-quartet of resonances with orbital angular momentum $L=2$ and
spin $S=3/2$ coupling to $J=1/2,\cdots,7/2$. Just below 1.9\,GeV we
find a spin doublet of resonances with $J^P=1/2^-$ and $3/2^-$.
Since a spin partner with $J^P=5/2^-$ is missing at this mass, the
two resonances form a spin doublet which must have a symmetric
orbital-angular-momentum wave function with $L=1$. For another set
of solutions, the four positive-parity resonances are accompanied by
mass-degenerate negative-parity partners -- as suggested by the
conjecture of chiral symmetry restoration. The possibility of a
$J^P=1/2^+, 3/2^+$ spin doublet at 1900\,MeV belonging to a 20-plet
is discussed.
} 
\maketitle

\mail{klempt@hiskp.uni-bonn.de\\
The $\pi N$ induced amplitudes, photoproduction observables and
multipoles for both solutions (BG2011-01 and 02) can be downloaded
from our web site as figures or in the numerical form
(http://pwa.hiskp.uni-bonn.de).}

\maketitle

\section{Introduction}
One of the most striking successes of the quark model was the
successful calculation of the spectrum and of the mixing angles of
low-lying excited baryons, using a harmonic oscillator potential and
a hyperfine interaction as derived from first-order perturbative QCD
\cite{Isgur:1977ef,Isgur:1978xj,Isgur:1978wd}. In the course of
time, the model was improved by taking into account relativistic
corrections \cite{Capstick:1986bm} or by fully relativistic
calculations \cite{Loring:2001kx}. The effective one-gluon exchange
interaction
\cite{Isgur:1977ef,Isgur:1978xj,Isgur:1978wd,Capstick:1986bm} was
replaced by one-boson exchange interactions between constituent
quarks \cite{Glozman:1997ag} or by in\-stanton-indu\-ced
interactions \cite{Loring:2001kx}. The improved baryon wave
functions provided the basis for calculations of partial decay
widths
\cite{Capstick:1992th,Capstick:1993kb,Capstick:1998uh,Melde:2005hy,Migura:2006en},
of helicity amplitudes \cite{Capstick:1992uc,VanCauteren:2005sm},
and of form factors
\cite{Glozman:2001zc,Berger:2004yi,VanCauteren:2003hn,Melde:2007zz}.
The spectrum of nucleon and $\Delta$ resonances calculated on the
lattice \cite{Edwards:2011jj} shows striking agreement with
expectations based on quark models: with increasing mass, there are
alternating clusters of states with positive and with negative
parity. No parity doubling is observed, and the number of expected
states on the lattice and in quark models seems to be the same.
 In spite of these considerable successes, there are
still many open questions in baryon spectroscopy (for reviews, see
\cite{Hey:1982aj,Capstick:2000qj,Klempt:2009pi}) which need further
investigations.

i. There are some low-mass baryon resonances which resist a
straightforward interpretation as quark model states. In effective
field theories of strong interactions, these resonances can be
generated dynamically from meson-baryon interactions. Well-known
examples are the Roper resonance $N_{1/2}^+(1440)$
\cite{Krehl:1999km}, the $N_{1/2^-}(1535)$ \cite{Kaiser:1995cy}, and
the $\Lambda_{1/2^-}(1405)$ \cite{Jido:2003cb}. The relation between
quark-model states and those dynamically generated is not yet
explored.

ii. There is the problem of the \emph{missing resonances}: quark
models predict many more resonances than have been observed so far,
especially at higher energies. This problem is aggravated by the
prediction of additional states, hybrid baryons, in which the
gluonic string mediating the interaction between the quarks is
itself excited \cite{Capstick:2002wm}. Their spectrum is calculated
to intrude the spectrum of baryon resonances at 2\,GeV and above.

iii. Baryon excitations come very often in parity doublets, of pairs
of resonances having opposite parity but similar mas\-ses. The
occurrence of parity doublets is unexpected in quark models. It has
led to the conjecture that chiral symmetry could be restored when
baryons are excited
\cite{Glozman:1999tk,Jaffe:2004ph,Glozman:2007ek,Glozman:2008vg}. It
has been suggested that the transition from constituent quarks to
current quarks can be followed by precise measurements of the masses
of excited hadron resonances \cite{Bicudo:2009cr}. The splitting in
(squared) mass between two states forming a parity doublet (like
$\Delta(1950)$ $F_{17}$ and $\Delta(2200)G_{17}$, 1.04\,GeV$^2$) is
slightly smaller than the mean mass square difference per unit of
angular momentum (the string tension, 1.1\,GeV$^2$). This effect can
possibly be interpreted as weak attraction between parity partners
in the 2\,GeV mass region and as onset of a regime in which chiral
symmetry is restored \cite{Klempt:2002tt}. This possibility depends,
of course, crucially on the assumption that chiral symmetry is
restored in the high-mass part of the hadron excitation spectrum.

iv. A very simple phenomenological two-parameter mass formula
\cite{Klempt:2002vp} describes the baryon mass spectrum with an
unexpectedly good precision. The formula can been derived
\cite{Forkel:2008un} within an analytically solvable
``gravitatio\-nal" theory simulating QCD \cite{Brodsky:2010kn} which
is defined in a five-di\-men\-sional Anti-de Sitter (AdS) space
embedded in six dimensions. The mass formula predicts
mass-degenerate spin multiplets with defined orbital angular
momentum $L$ in which the orientation of the quark spin $S$ relative
to $L$ has no or little impact on the baryon mass, a prediction
which needs to be confirmed or rejected by further experimental
information.

\begin{figure*}[pt]
\begin{tabular}{cccc}
             \epsfig{file=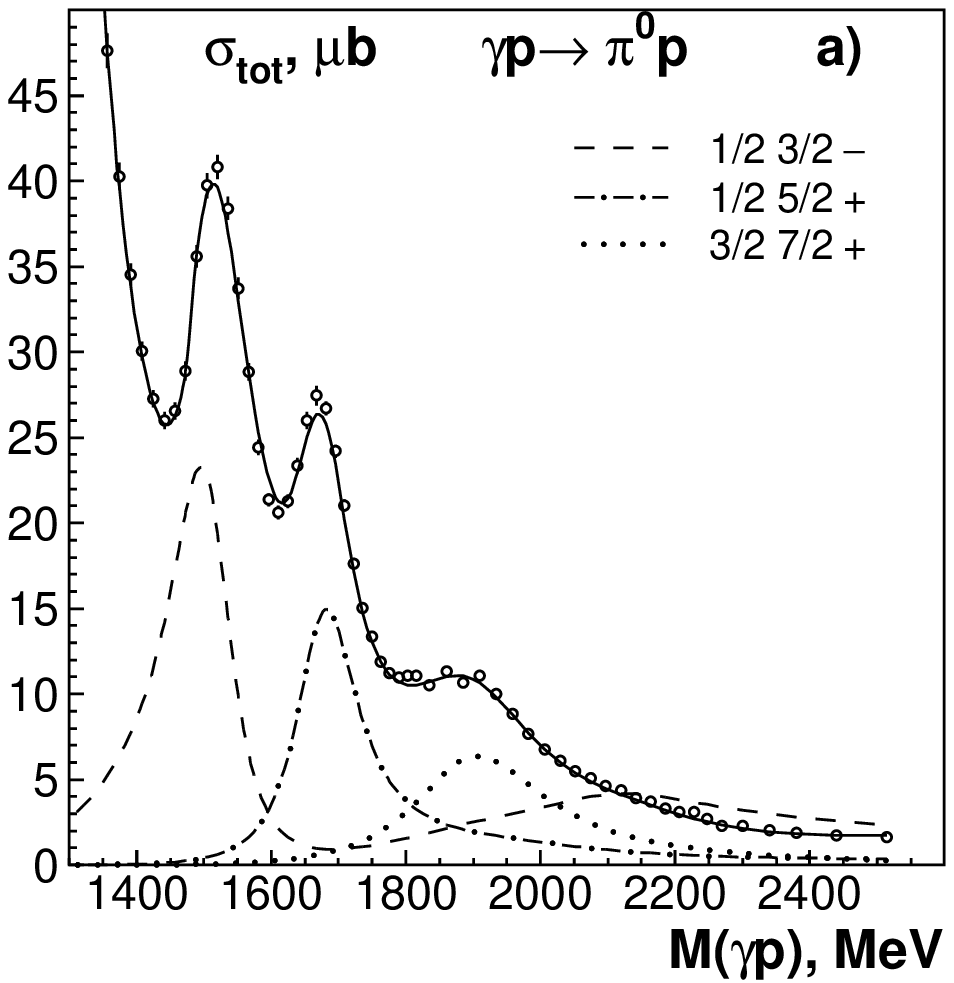,width=0.24\textwidth}&
\hspace{-3mm}\epsfig{file=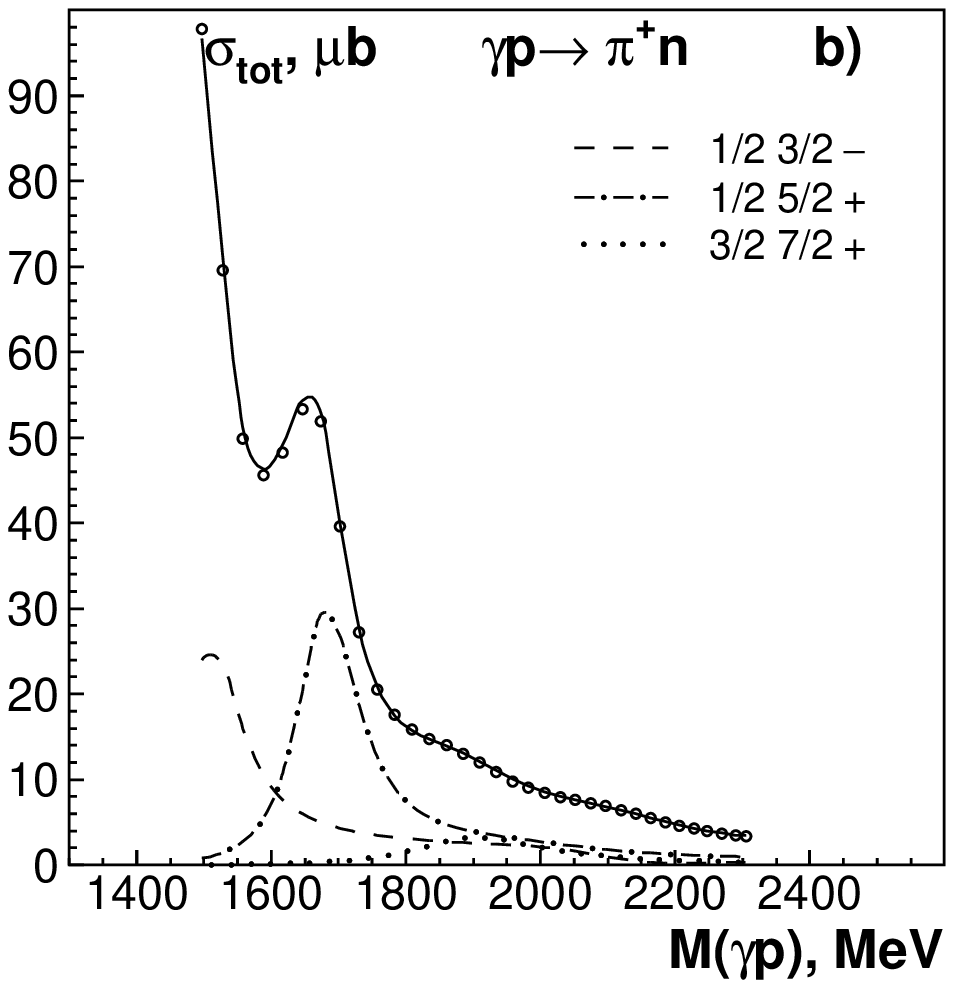,width=0.24\textwidth}&
\hspace{-3mm}\epsfig{file=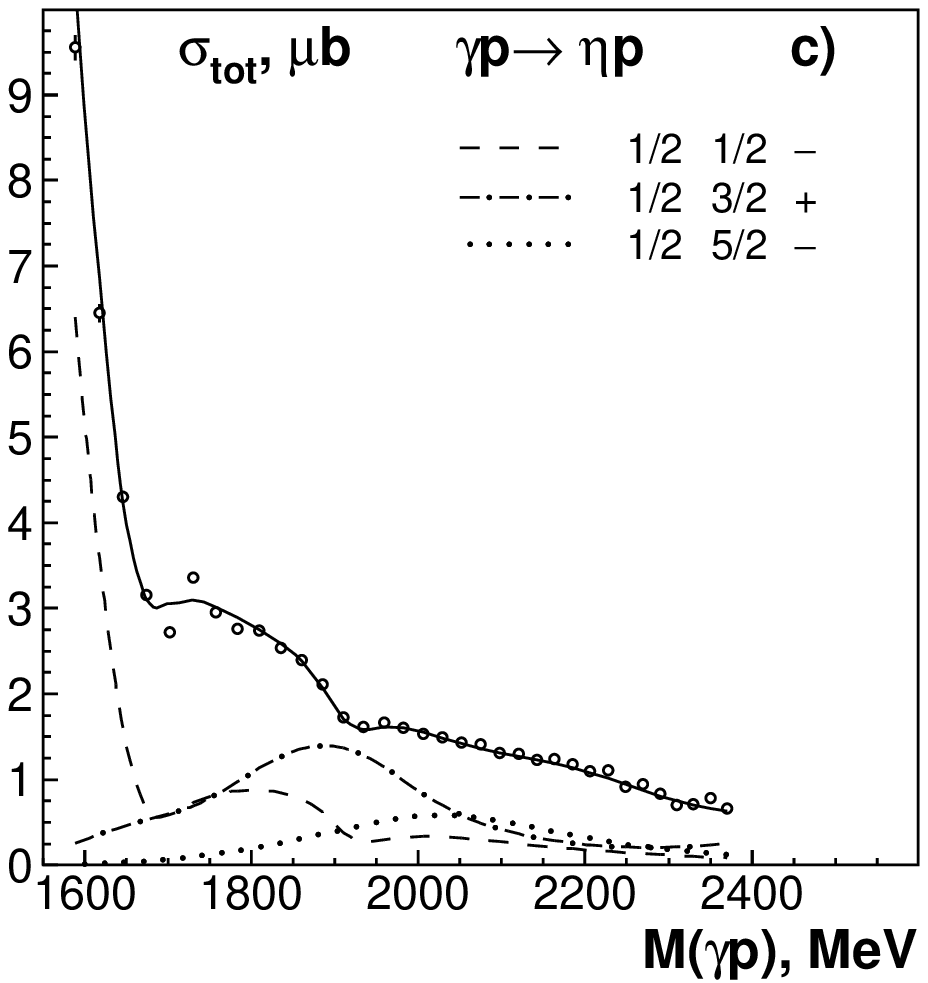,width=0.24\textwidth}&
\hspace{-3mm}\epsfig{file=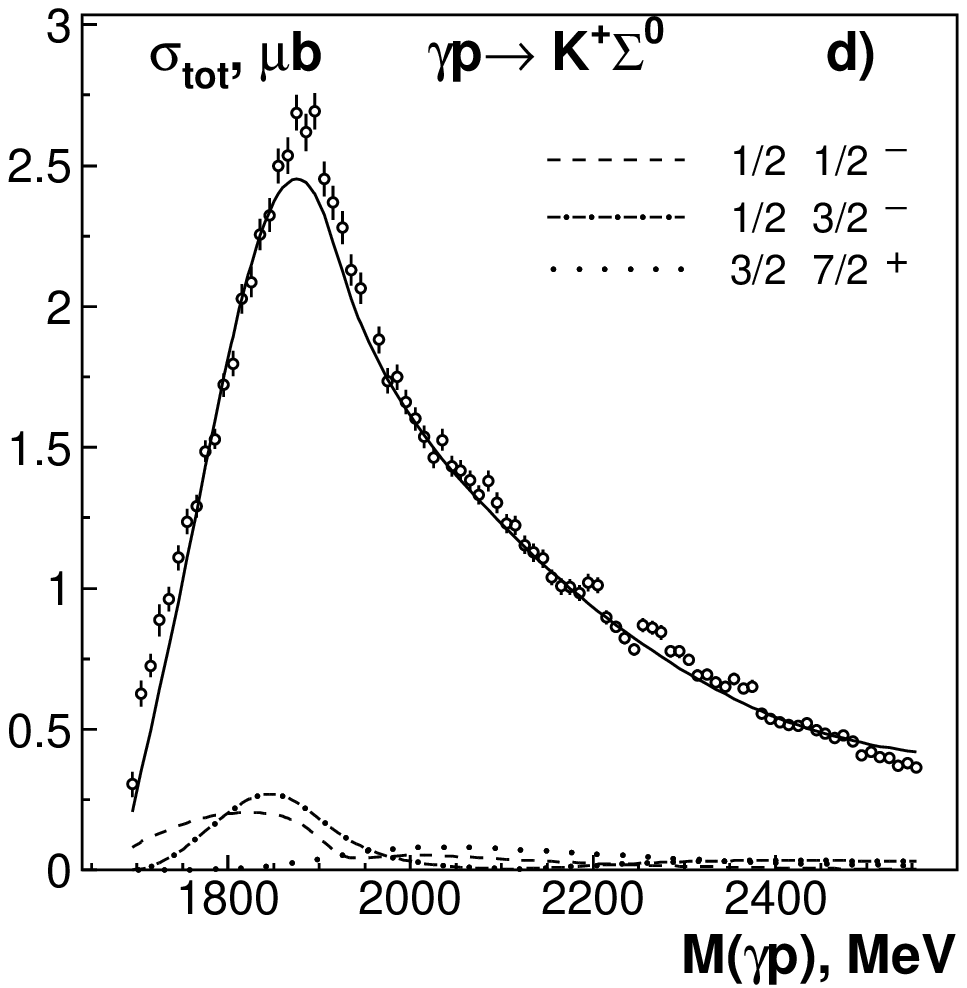,width=0.24\textwidth}
\end{tabular}
\\
\begin{minipage}[c]{0.72\textwidth}
\begin{tabular}{ccc}
             \epsfig{file=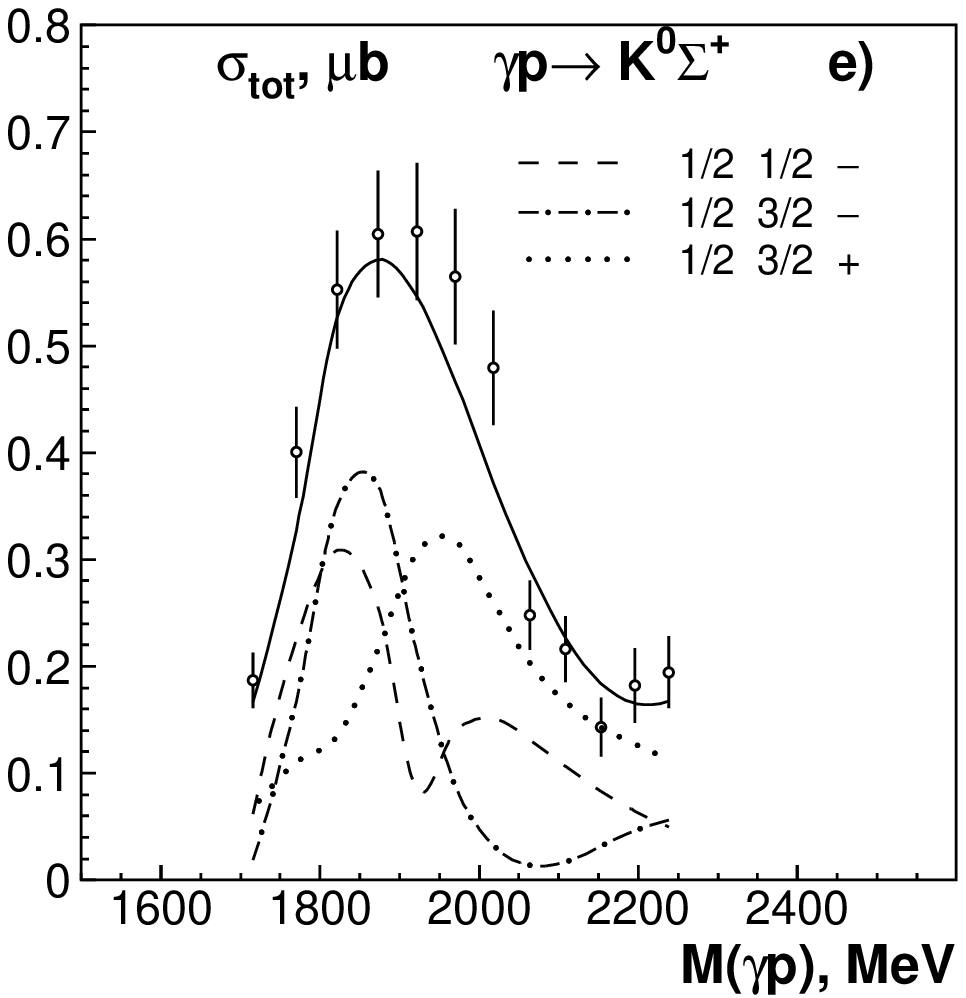,width=0.333\textwidth}&
\hspace{-3mm}\epsfig{file=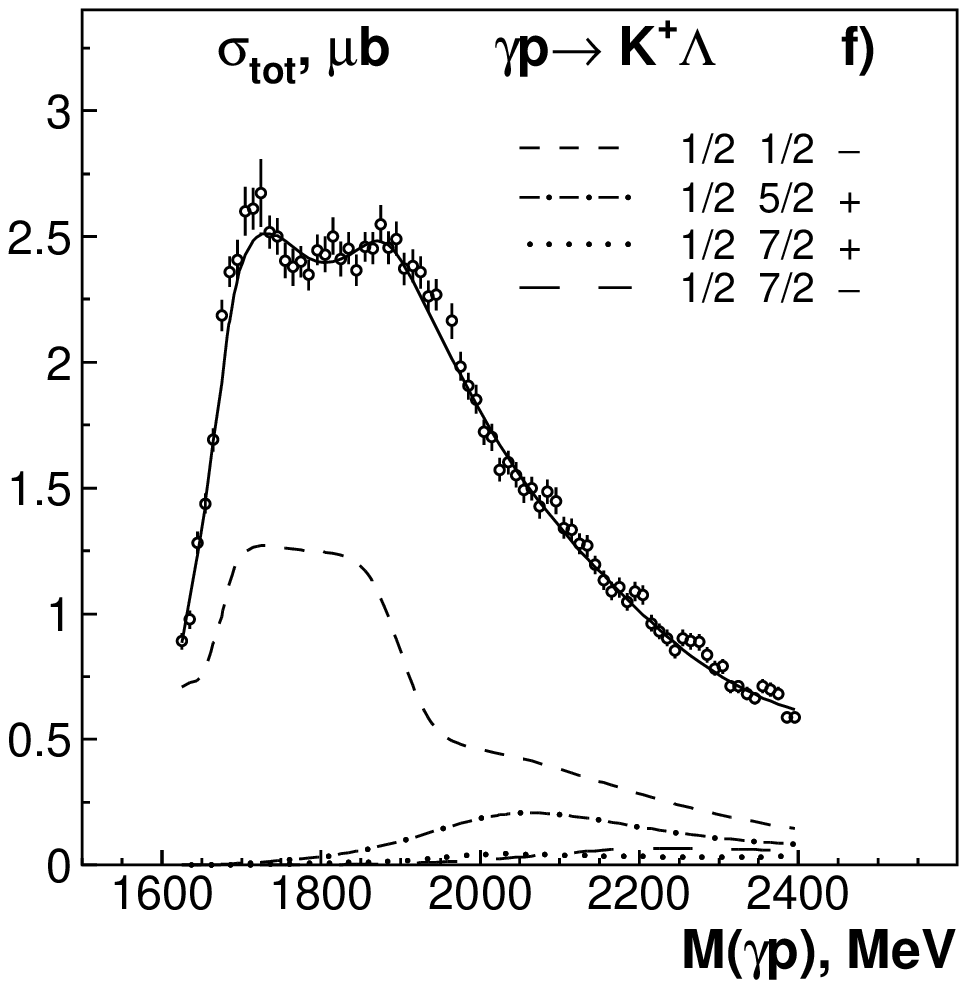,width=0.333\textwidth}&
\hspace{-3mm}\epsfig{file=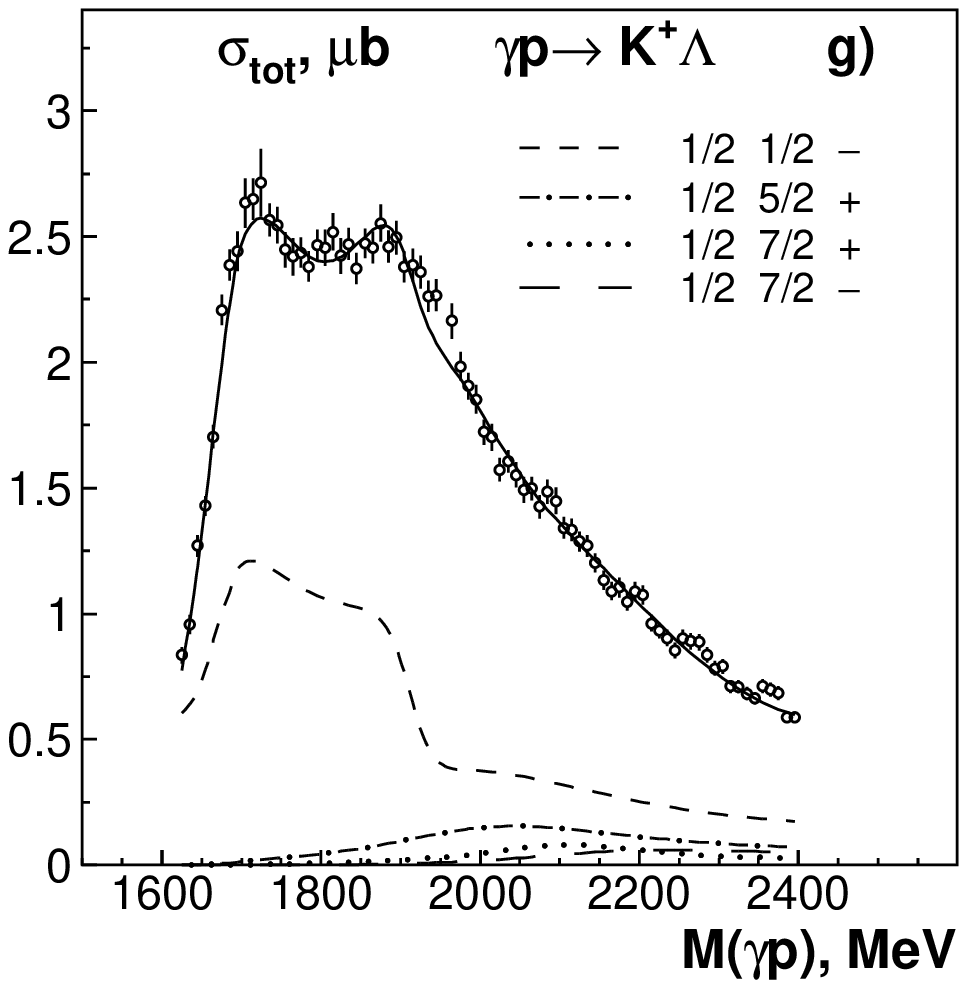,width=0.333\textwidth}
\end{tabular}
\end{minipage}
\hspace{7mm}\begin{minipage}[c]{0.23\textwidth}
\caption{\label{sig_tot} The total cross secti\-ons determined from
the data (see text for the details) a)
\cite{Bartholomy:2004uz,vanPee:2007tw}, b) \cite{Dugger:2009pn}, c)
\cite{Crede:2009zzb}, d) \cite{Dey:2010hh}, e) \cite{Ewald:2011a},
and f,g) \cite{McCracken:2009ra}, and contributions from partial
waves with resonances in the fourth resonance region (solution
BG2011-02). In (f) and (g), the contributions from two solutions
labeled BG2011-01 and BG2011-02 are shown.}
\end{minipage}\vspace{-4mm}
\end{figure*}

In this paper, we report properties of nucleon resonances with
masses of about 2\,GeV, a mass range which is often called 4$^{\rm
th}$ resonance region. The resonance regions are well seen in Fig.
\ref{sig_tot}a, starting with the $\Delta(1232)$ tail as 1$^{\rm
st}$ resonance region followed by peaks,  indicating the 2$^{\rm
nd}$, and 3$^{\rm rd}$ and 4$^{\rm th}$ resonance region. The
properties of nucleon resonances are determined in a coupled-channel
partial wave analysis of a large body of data. The analysis methods
are documented in
\cite{Anisovich:2004zz,Klempt:2006sa,Anisovich:2006bc,Anisovich:2007zz},
earlier results can be found in \cite{Anisovich:2005tf,Sarantsev:2005tg,%
Anisovich:2007bq,Nikonov:2007br,Anisovich:2008wd,Anisovich:2009zy}.
Recently, we have enlarged considerably our data base. In
particular both, the recent high-statistics data \cite{McNabb:2003nf,%
Zegers:2003ux,Lawall:2005np,Bradford:2005pt,Bradford:2006ba,Lleres:2007tx,%
Castelijns:2007qt,Lleres:2008em,McCracken:2009ra} on photoproduction
of hyperons but also the old low-statistics data on pion-induced
hyperon production
\cite{Knasel:1975rr,Baker:1978qm,Saxon:1979xu,Hart:1979jx,Candlin:1982yv}
proved to be very sensitive to properties of contributing
resonances. In the first paper \cite{Anisovich:2010an}, we reported
evidence for $N(1710)P_{11}$, $N(1875)P_{11}$, $N(1900)P_{13}$,
$\Delta(1910)P_{31}$, $\Delta(1600)P_{33}$ and $\Delta(1920)P_{33}$.
The main aim was to investigate if resonances seen in the
Karlsruhe-Helsinki (KH84) \cite{Hohler:1984ux} and the
Carnegie-Mellon (CM) \cite{Cutkosky:1980rh} analyses of $\pi N$
elastic scattering but not observed by the Data Analysis Center at
GWU \cite{Arndt:2006bf} can be identified in inelastic reactions.
The positive answer encourages us to study the ``full"` spectrum of
nucleon resonances in the 2\,GeV region.

The data on hyperon production with the largest statistics are those
with $K\Lambda$ in the final state. These are sensitive to nucleon
resonances only. Data on $K\Sigma$ contribute to both, to nucleon
and $\Delta$ resonances, the statistical errors for data on this
channel are considerably higher. For this reason, we concentrate
here on nucleon resonances. $\Delta$ resonances will be discussed
when new high-statistics data on the $\gamma p\to \pi^0\pi^0 p$ and
$p\pi^0\eta$ production channels from ELSA are included in our data
base.

At present, existence and properties of nucleon resonances are
mostly derived from energy dependent fits to the energy-independent
partial wave analyses of $\pi N$ elastic scattering data
\cite{Hohler:1984ux,Cutkosky:1980rh,Arndt:2006bf}. Large
discrepancies between the results of the three groups reveal the
weak points of these analyses: first, the amplitudes cannot be
constructed from the data without theoretical input. For a full
amplitude reconstruction, the differential cross section
$d\sigma/d\Omega$, target asymmetry, and spin rotation parameters
need to be known in the entire energy and angular range. Without
spin rotation parameters, only the absolute values of the spin-flip
and spin non-flip amplitudes $|H|$ and $|G|$ can be determined but
not their phases. Dispersion relations must be enforced but,
unfortunately, their impact on the partial wave solutions seems not
very well defined (judged from the differences in the results
obtained by \cite{Hohler:1984ux,Cutkosky:1980rh,Arndt:2006bf}).
Second, the inelasticity of the $\pi N$ amplitude is a free fit
parameter to be determined for every bin. It is unconstrained by the
decay modes of resonances. In practice, this freedom can be
exploited to fit the elastic partial wave amplitudes with equally
acceptable $\chi^2$ using a variety of models with a different
content of resonances. In the approach presented here, the
inelasticity of the energy independent elastic $\pi N$ amplitudes
are constrained by a large number of data sets on inelastic
reactions. The inelasticity of the $\pi N$ amplitude is no longer a
free fit parameter to be determined for every bin. Thresholds,
couplings to the different channels, and the opening of new channels
are properly taken into account.

We use the naming scheme adopted in \cite{Anisovich:2010an}. For
resonances listed in Review of Particle Properties (RPP)
\cite{Nakamura:2010zzi} we use the conventional names of the
Particle Data Group: $N({\rm mass})L_{2I,2J}$ and $\Delta({\rm
mass})L_{2I,2J}$ where $I$ and $J$ are isospin and total spin of the
resonance and $L$ the orbital angular momentum in the decay of the
resonance into nucleon and pion. For resonances not included in
\cite{Nakamura:2010zzi}, we use $N_{J^P}({\rm mass})$ and
$\Delta_{J^P}({\rm mass})$ which gives the spin-parity of the
resonance directly.

\section{\label{method}Data, PWA method, and fits}

The coupled-channel partial wave analysis uses the $\pi N$ elastic
amplitudes, alternatively from KH84 \cite{Hohler:1984ux} or from
SAID \cite{Arndt:2006bf}. SAID results are given with errors; for
KH84, no errors are given. We assume $\pm 5$\% errors for the fits.
Data are included on the reactions $\pi^-p\to \eta n$, $\pi N\to
K\Lambda$, $\pi N\to K\Sigma$, $\gamma p\to \pi^0 p$, $\gamma p\to
\pi^+ n$,$\gamma p\to K\Lambda$, $\gamma p\to K\Sigma$, $\pi^-p\to
\pi^0\pi^0n$, and $\gamma p\to \pi^0\pi^0 p$, $\gamma p\to \pi^0\eta
p$. Measurements of polarization variables, with polarization in the
initial or final state or with target polarization, are included in
the ana\-lysis whenever such data are available. A complete list of
the reactions and references to the data is given in Tables 1-5 of
reference \cite{Anisovich:2010an}. The data sets we added here are
differential cross section and recoil asymmetry on the $\pi^+ p\to
K^+\Sigma^+$ reaction in the mass region below 1850 MeV
\cite{Winik:1977mm,Crawford:1962zz,Baltay:1961,Carayannopoulos:1965,Bellamy:1972fa},
new CLAS data on $\gamma p -> K^+ \Sigma^0$ \cite{Dey:2010hh}, new
MAMI data on $\gamma p\to \eta p$ \cite{McNicoll:2010qk} and for the
reaction $\gamma p\to K^0\Sigma^+$ the differential cross section
\cite{Ewald:2011a}. We now excluded from the analysis the $\gamma
p\to \eta p$ data on the target asymmetry: since there seems to be
an inconsistency between these data and new preliminary CB-ELSA/TAPS
data \cite{Hartman}.

The MAMI data were well fitted, without need for significant changes
in mass, widths or coupling constants of the contributing
resonances. In a narrow mass region at about 1700\,MeV,
statistically significant deviations between data and fit showed up.
Dedicated fits were made \cite{Anisovich:2011ka} to study this
effect. The fits improved by introduction of a narrow resonance, or
by introducing a $\omega N$ coupling in the $S_{11}$ wave.
Suggestions for experiments were made capable to decide which
alternative is realized in nature. For most resonances, the ELSA
data set had no noticeable effect on their properties neither. But
when these data were introduced in the fit, an additional resonance
$N_{3/2^-}(1870)$ was needed (which was then also of substantial
help in describing other data sets).

\begin{figure}[pt]
\bc
\hspace{-2mm}\epsfig{file=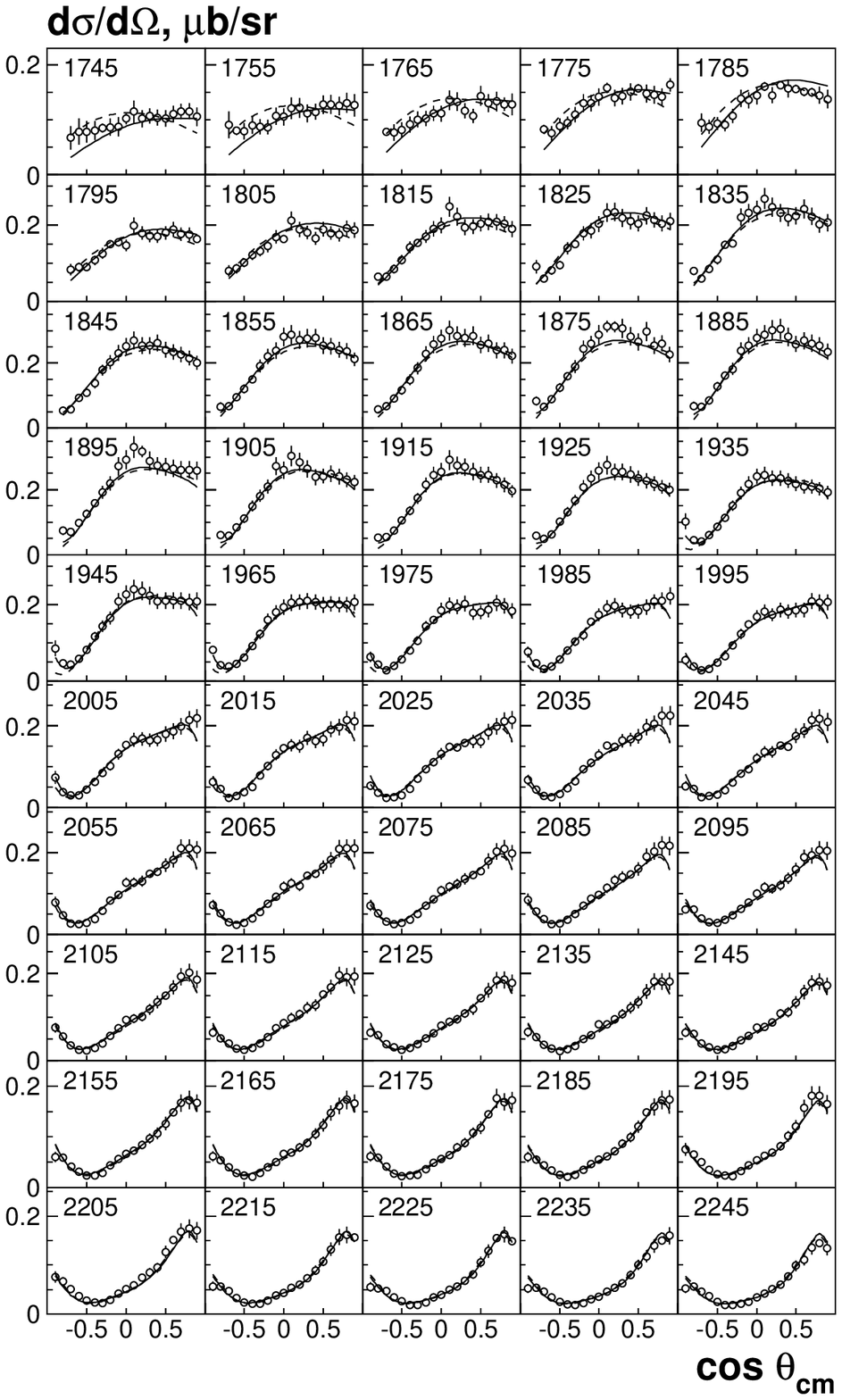,height=0.51\textheight,clip=on}
\vspace{-3mm}
 \ec
\caption{\label{gp_ksig_dcs}Differential cross section for $\gamma
p\to K^+\Sigma^0$ \protect{\cite{Dey:2010hh}}. The full curve
corresponds to solution BG2011-02 ($\chi^2$=1.44), the dashed one to
BG2011-01 ($\chi^2$=1.44).}
\bc
\hspace{-2mm}\epsfig{file=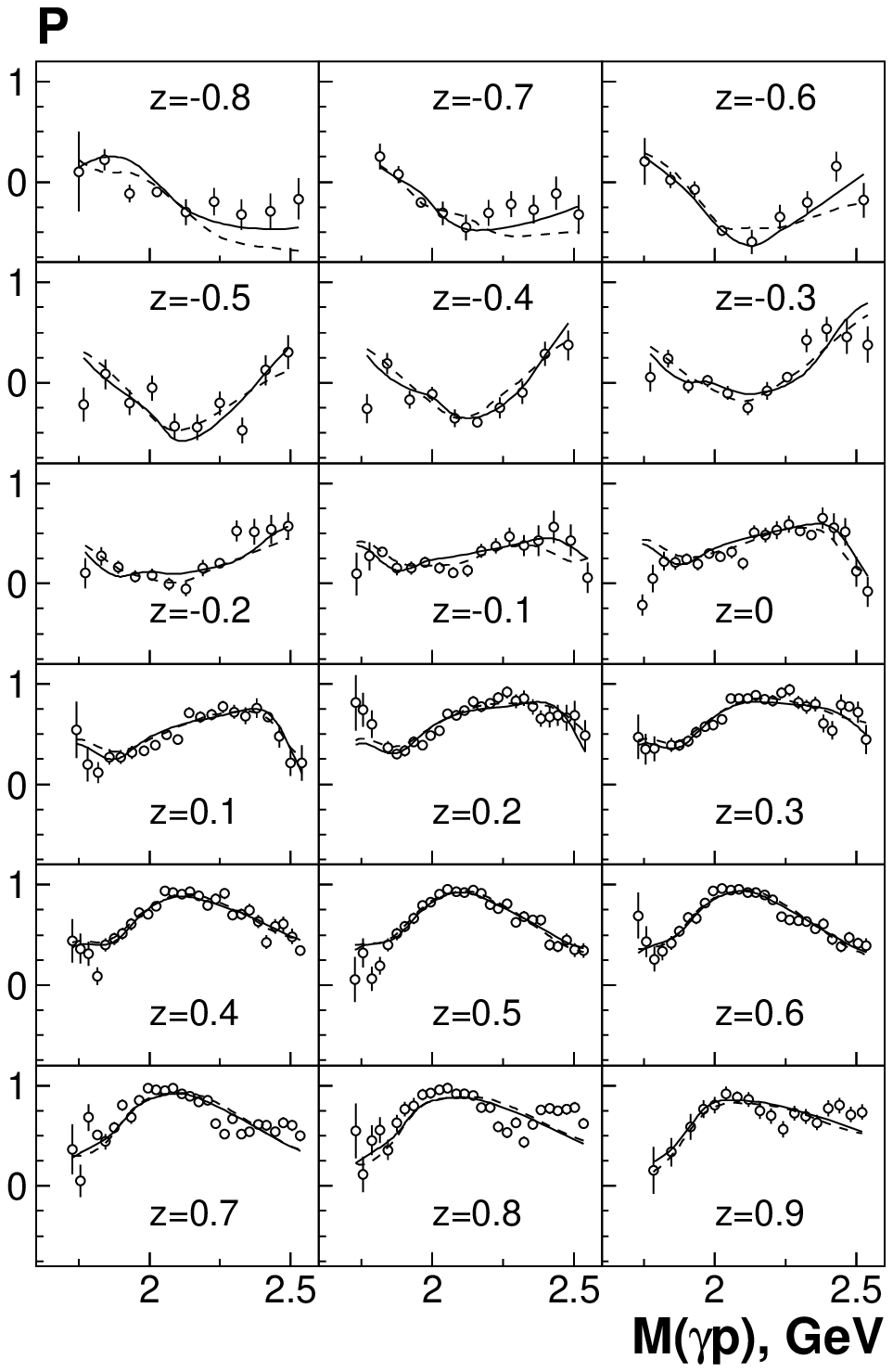,height=0.37\textheight,clip=on}
\vspace{-3mm}
\ec
\caption{\label{gp_ksig_p}Recoil polarization for $\gamma p\to
K^+\Sigma^0$ \protect{\cite{Dey:2010hh}}. The full curve corresponds
to solution BG2011-02 ($\chi^2$=2.70), the dashed one to BG2011-01
($\chi^2$=2.76).}
\end{figure}
\begin{figure}[h]
\hspace{-2mm}\epsfig{file=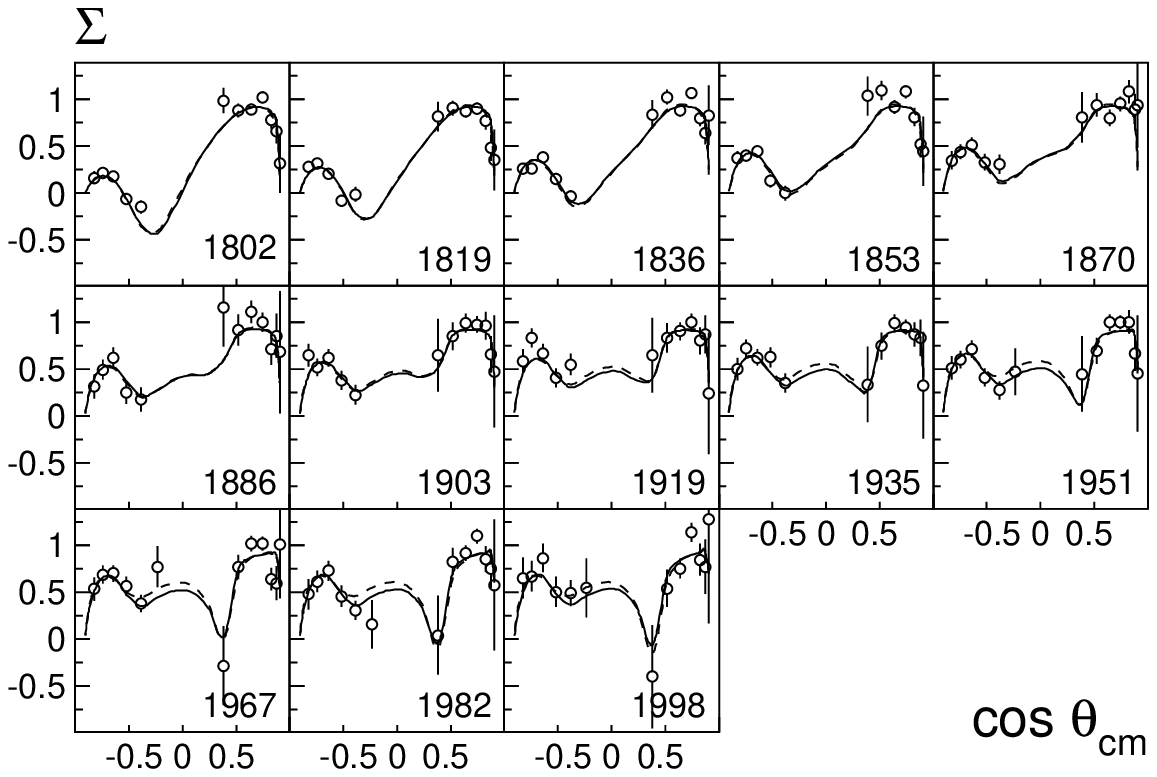,width=0.49\textwidth}
\caption{\label{gp_sigma_p} Beam asymmetry for single-$\pi^0$
photoproduction $\gamma p\to \pi^0 p$ reaction \cite{Sparks:2010vb}.
The full curve corresponds to the solution BG2011-02 and the dashed
one corresponds to the solution BG2011-01.}
\begin{tabular}{cc}
\hspace{-2mm}\epsfig{file=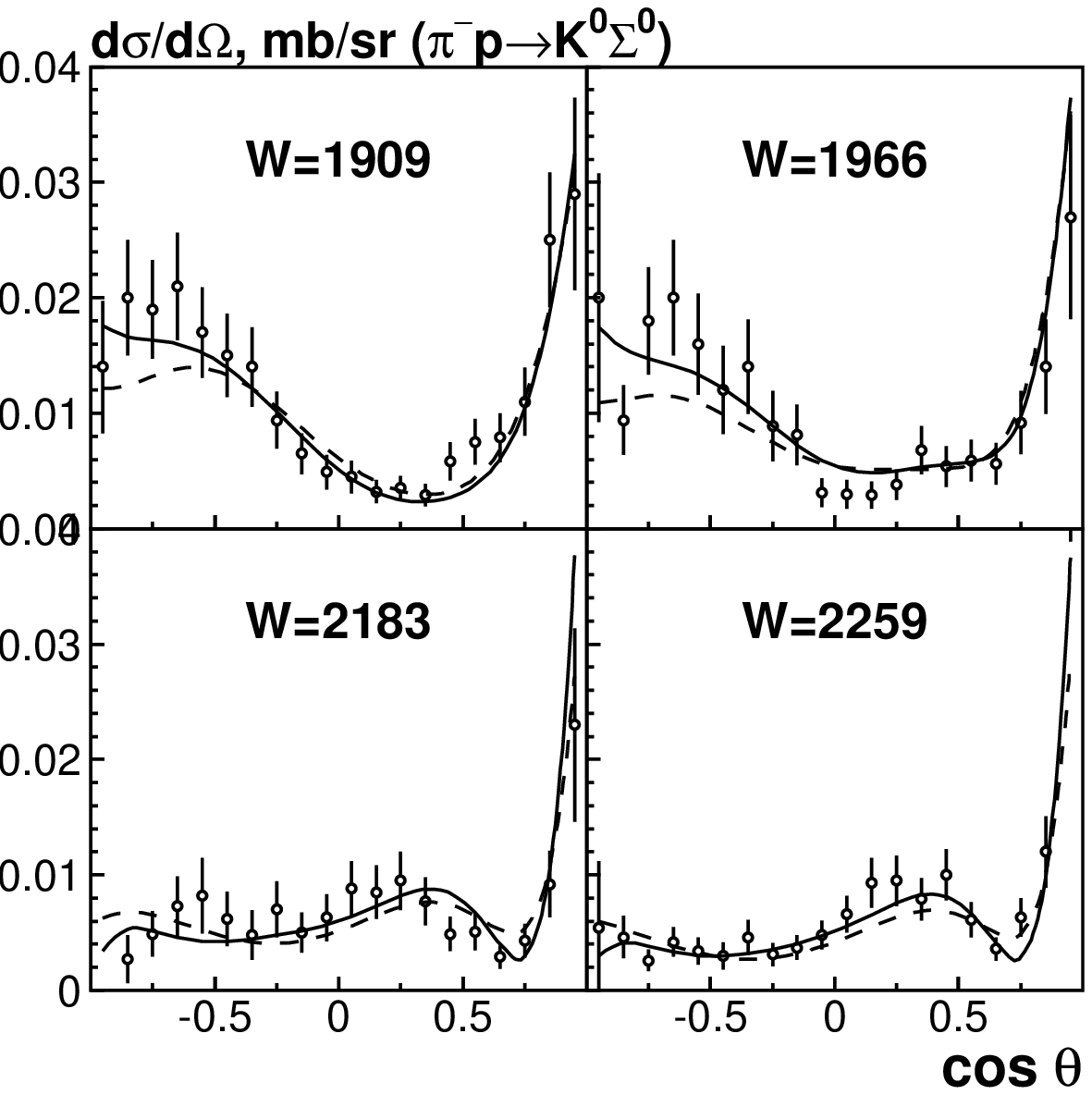,width=0.24\textwidth}&
\hspace{-2mm}\epsfig{file=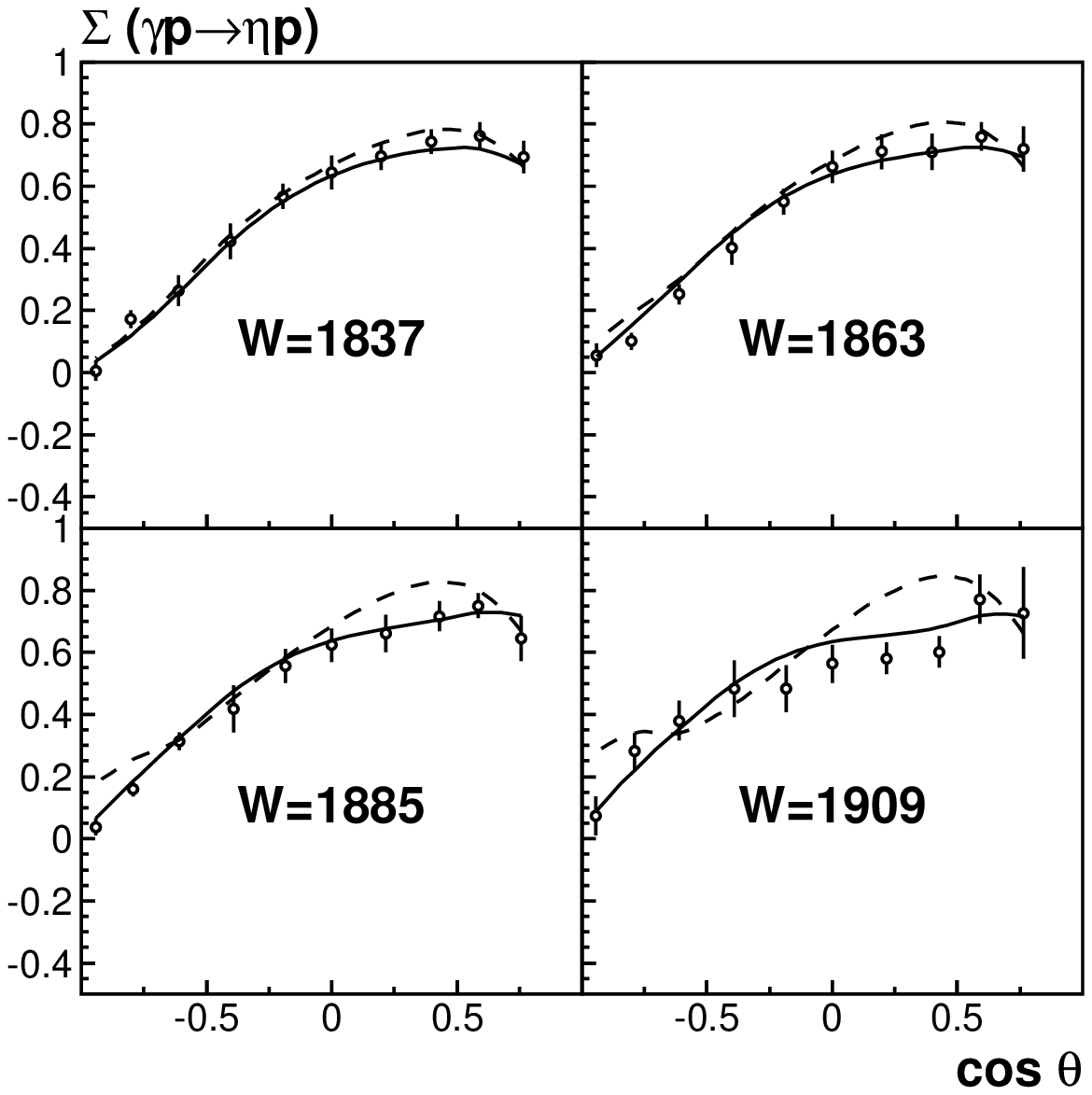,width=0.24\textwidth}
\end{tabular}
\caption{\label{sigma_pio}Left: Differential cross section for
$\pi^- p\to K^0\Sigma^0 $ \cite{Hart:1979jx}. Right: Beam asymmetry
from GRAAL for $\gamma p\to \eta p$ \cite{Bartalini:2007fg}. The
full curves correspond to the solution BG2011-02, the dashed ones to
the best fit without a second $5/2^-$ resonance. \vspace{2mm}}
\begin{tabular}{cc}
\hspace{-2mm}\epsfig{file=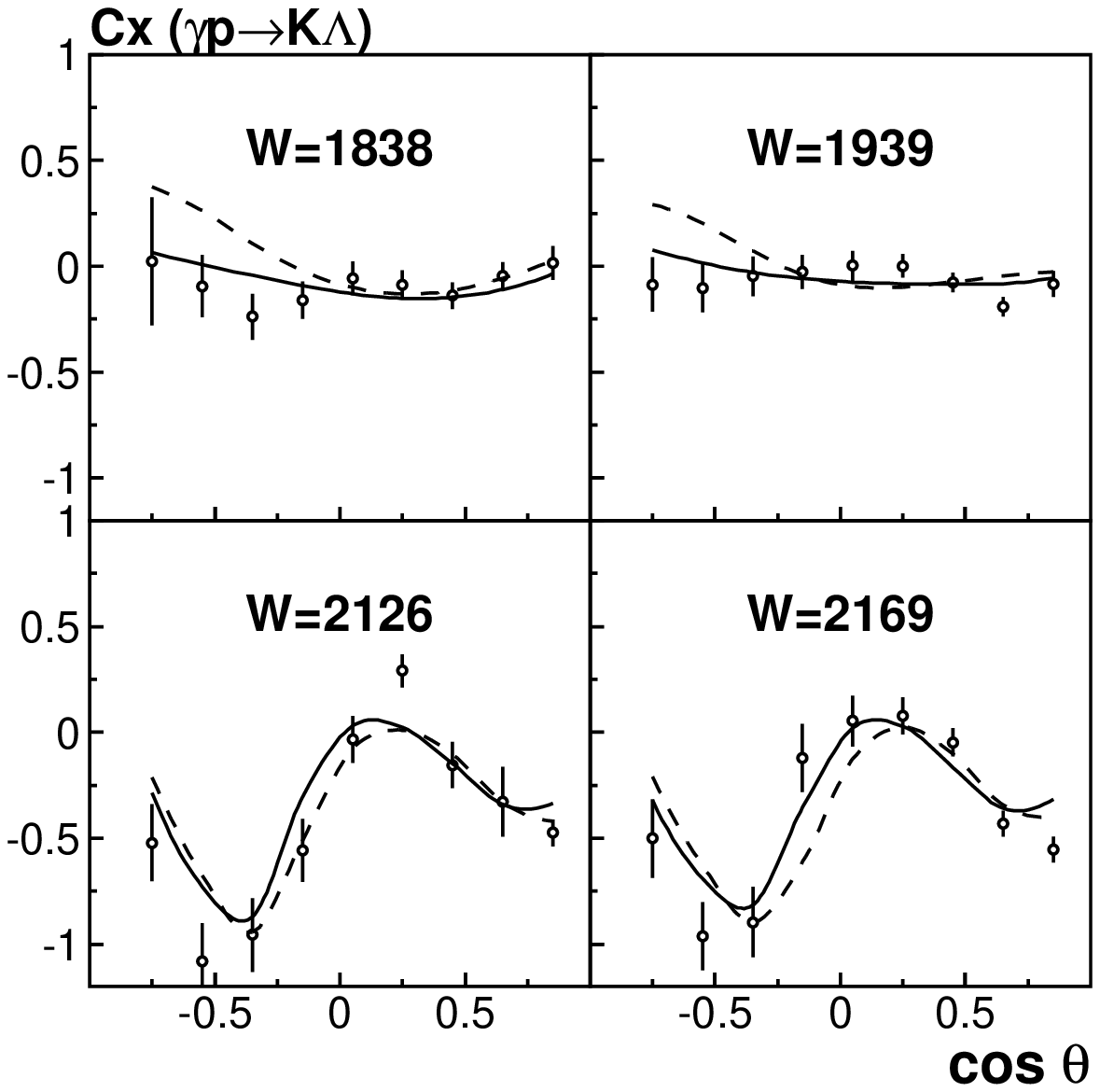,width=0.24\textwidth}&
\hspace{-2mm}\epsfig{file=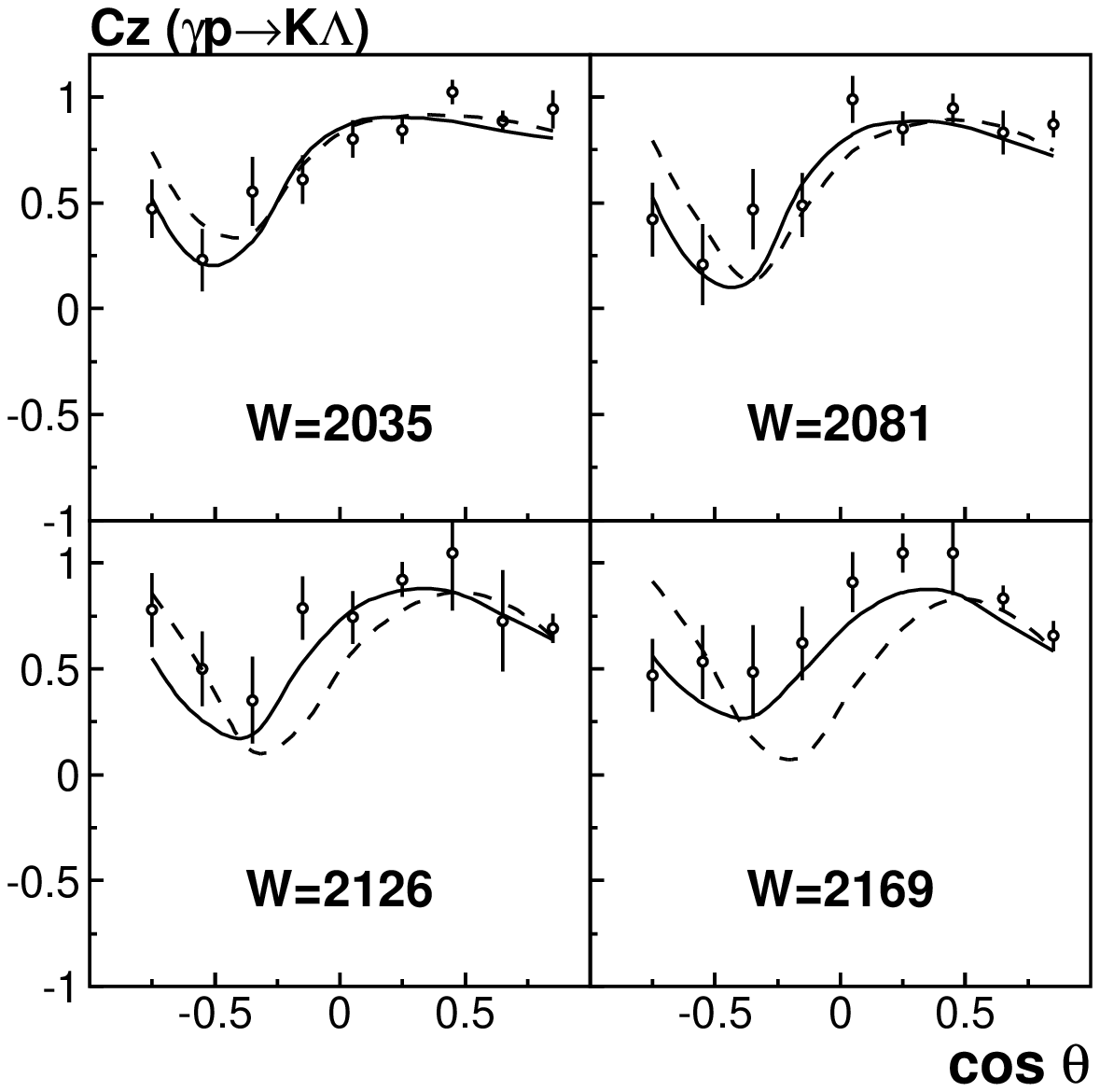,width=0.24\textwidth}
\end{tabular}
\caption{\label{pol_c}Polarization transfer variables $C_x$, $C_z$
from circularly polarized photons to $\Lambda$ hyperons
\cite{Bradford:2006ba}. The full curves correspond to the solution
BG2011-02, the dashed ones to the best fit without a resonant
contributions in the nucleon $5/2^+$ partial wave above 1.8
GeV.}\vspace{-3mm}
\end{figure}

The analysis method
\cite{Anisovich:2004zz,Klempt:2006sa,Anisovich:2006bc,Anisovich:2007zz},
uses relativistic invariant operators which are constructed directly
from the 4-vectors of the particles. Resonances are parameterized in
multi-channel K matrices. Background terms are partly added within
the K-matrix either as constants or in the form $(a+b\sqrt
s)/(s-s_0)$ with $s_0$ simulating left-hand cuts ($s_0$ being
negative, at a few GeV$^2$), partly they are added as $t$-channel
meson-exchange or $u$-channel baryon-exchange amplitude. These
amplitudes are sufficiently flexible to describe the data mentioned
above with good precision.  In the pion induced reactions the $t-$
and u-channel exchanges are projected into the partial waves and
contribution from lowest partial waves (up to spin 3/2) are
subtracted from the t and u-exchange amplitudes. Thus, for the
lowest partial waves, these exchanges are taken effectively into
account as K-matrix non-resonant terms. Such approach ensures that
we fully satisfy the unitarity condition for the pion induced
reactions. We remind the reader that we found two classes of
solutions, called BG2010-01 and BG2010-02. In this article, figures
are shown for solutions which are called BG2011-01 and BG2011-02
which are modifications of solutions BG2010-01 and BG2010-02 due to
including/excluding of the data sets discussed above and including a
number of high mass poles into the K-matrix parameterization.
Contributions of partial waves with resonances in the fourth
resonance region are shown in Fig.~\ref{sig_tot}. The total cross
sections are determined by summation over the (binned) experimental
differential cross sections, including statistical and systematic
errors, and using our partial wave results in the angular range
where no data exist. In Figs.~\ref{gp_ksig_dcs} to \ref{pol_c} we
exemplify the fit quality by showing a few data sets.

As can be seen from the figures, there are only marginal differences
in fit quality between the two solutions. Indeed, the overall
$\chi^2$ of the two fits hardly differs. Many resonances have very
similar properties in both types of solutions but some are
distinctively different. Significant differences are found, e.g., in
the helicity amplitudes for the production of $N(1710)P_{11}$ and
$N(1720)P_{13}$, their $N\eta$ decay fractions
\cite{Anisovich:2010an} and in the masses and widths of a few
resonances with low RPP star-rating. Within both classes of
solutions, a large variety of different fits were made, e.g. by
adding a further pole in a particular partial wave or by changing
start values of the fit. The spread of results within a class of
solutions is used to define the errors. The statistical errors
returned by the fit are usually unrealistically small.

In Figs. \ref{sigma_pio} and \ref{pol_c} we show a few examples,
data and fit curves, covering the central part of the fourth
resonance region. In addition to the main fit represented by a solid
curve we also show fits (dashed curve) in which one of the
resonances is removed. These curves are discussed below. The fits
are meant to illustrate the significance of a particular data set to
resonance formation. Further plots can be seen in
\cite{Anisovich:2010an} and on our web page
(http://pwa.hiskp.uni-bonn.de).

In the fits to the data, we use weighting factors $\omega$ to avoid
that low statistics polarization data are overruled by
high-statistics data on, e.g., differential cross sections. The
weights are adjusted to achieve a visually acceptable fit quality
for all data sets. The total $\chi^2$ is then calculated as weighted
sum of the $\chi^2$ from the individual data sets: $\sum \omega_i
N_i/\sum N_i$ where $\omega_i$ and $N_i$ is the weight and number of
events in the data set $i$. To the $\chi^2$ we add the likelihood
($\delta\chi^2 = 2\delta\ln{\cal L}$) of the event-based likelihood
fit to multi-body final states. The absolute value of the likelihood
depends on the normalization and has no meaning. Therefore only
differences of the total $\chi^2$ are given.

The fit quality decreases above 2.2 to 2.3\,GeV. We did not try to
systematically improve the description in this mass region. A large
number of resonances is expected, and the data base is certainly not
yet sufficient to arrive at solid conclusions.

\section{Discussion of partial waves}
\subsection{Positive parity nucleon resonances}
Positive-parity nucleon resonances above the nucleon may belong - in
the harmonic oscillator approximation - to the second excitation
band. Well known are the Roper resonance $N(1440)$ $P_{11}$, the 3-star
$N(1710)P_{11}$ (challenged by SAID \cite{Arndt:2006bf}), and the
spin-doublet $N(1720)P_{13}$ and $N(1680)F_{15}$. Above, there is a
possible quartet of nucleon resonances,

\vspace{2mm}
\centerline{
$N_{1/2+}(1875)$, $N(1900)P_{13}$, $N(2000)F_{15}$, $N(1990)F_{17}$.}
\vspace{2mm}

\noindent The $I(J^P)=\frac12(\frac12^+)$ resonance was reported in
\cite{Anisovich:2010an}, the other three resonances are listed in
RPP with two stars: evidence for their existence is fair. None of
them is needed in the SAID analysis \cite{Arndt:2006bf}. The next
known positive-parity nucleon resonance is the four-star
$N(2220)H_{19}$; it should have a spin partner with
$I(J^P)=\frac12(\frac72^+)$. These two resonances belong to fourth
excitation band.

\boldmath
\subsubsection{$I(J^P)=\frac12(\frac12) ^+$ and $\frac12(\frac32 ^+$)}
\unboldmath

Nucleon resonances in the $J^P=\frac12 ^+$ and $\frac32 ^+$ partial
waves have been discussed in \cite{Anisovich:2010an}. For
convenience of the reader, we recall the main features here.

Above the Roper resonance, we observe a pole, given as (Re$(pole),
-2{\rm Im}(pole)$), at ($1690^{+25}_{-10}, 210\pm25$)\,MeV or ($1695
\pm 15, 220\pm 30$)\,MeV  for BG2010-01 and BG2010-02, respectively,
which we identify with the known 3-star $N(1710)P_{11}$. A second
resonance is seen at ($1860\!\pm\!20, 110^{+30}_{-10}$)\,MeV  or
($1850^{+20}_{-50}, 360\!\pm\!40$)\,MeV, respectively, with a width
which depends sensitively on the solution. We call this resonance
$N_{1/2^+}$ $(1875)$ (this is the Breit-Wigner mass). A further pole
at (2100, 500)\,MeV improves the stability of the fit. The pole is
also reported by KH84 \cite{Hohler:1984ux} and CM
\cite{Cutkosky:1980rh}. We do neither claim its existence nor rule
it out.

The $I(J^P)=\frac12(\frac32) ^+$-wave houses, of course, the well
known $N(1720)P_{13}$. Its pole position is determined to ($1695\pm
30$, $400\pm 60$)\,MeV or ($1670\pm 30$, $420\pm 60$)\,MeV,
respectively. Above, there is at least one additional resonance
$N(1900)P_{13}$ \cite{Nikonov:2007br}; however, a better fit is
obtained when a two-pole structure is assumed. In
\cite{Anisovich:2010an}, we gave effective masses for the two
resonances. The effective mass and the effective width are defined
by the maximum and the full width at half maximum of the $\pi N\to
M\,B$ transition strength where $B,M$ stand for the final-state
baryon and meson. These effective values are more stable against
variations in the parameterization than Breit-Wig\-ner parameters:
the widths of resonances often become very large when three-body
final states are included. We find, as mean value from a few
transition amplitudes to different final states, ($M_{\rm eff},
\Gamma_{\rm eff}$) = ($1910\!\pm\!25$, $300\!\pm\!70$) and
($1970\!\pm\!20$, $220\!\pm\!60$)\,MeV for two resonances in the
$P_{13}$ wave. The two resonances differ in the photo-coupling;
their decay modes were rather similar. Hence doubts remain in the
existence of the upper resonance, in spite of the gain in $\chi^2$
when it is introduced. Future experiments will have to decide if the
splitting into two resonances is real. The existence of
$N(1900)P_{13}$ with ($M_{\rm eff}, \Gamma_{\rm eff}$) =
($1910\!\pm\!25$, $300\!\pm\!80$) is, however, mandatory to achieve
an acceptable fit.

\boldmath
\subsubsection{$I(J^P)=\frac12(\frac52^+)$}
\unboldmath

\begin{figure}[ht]
\epsfig{file=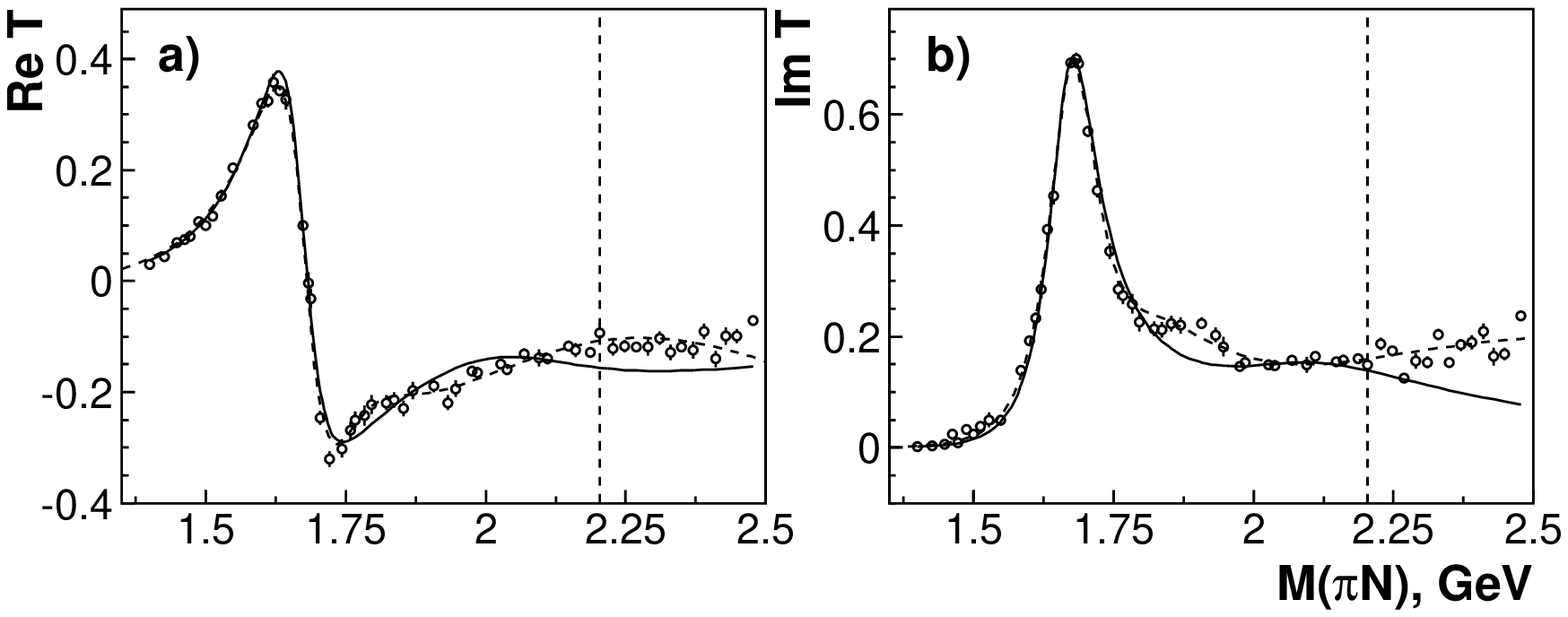,width=0.47\textwidth}
\epsfig{file=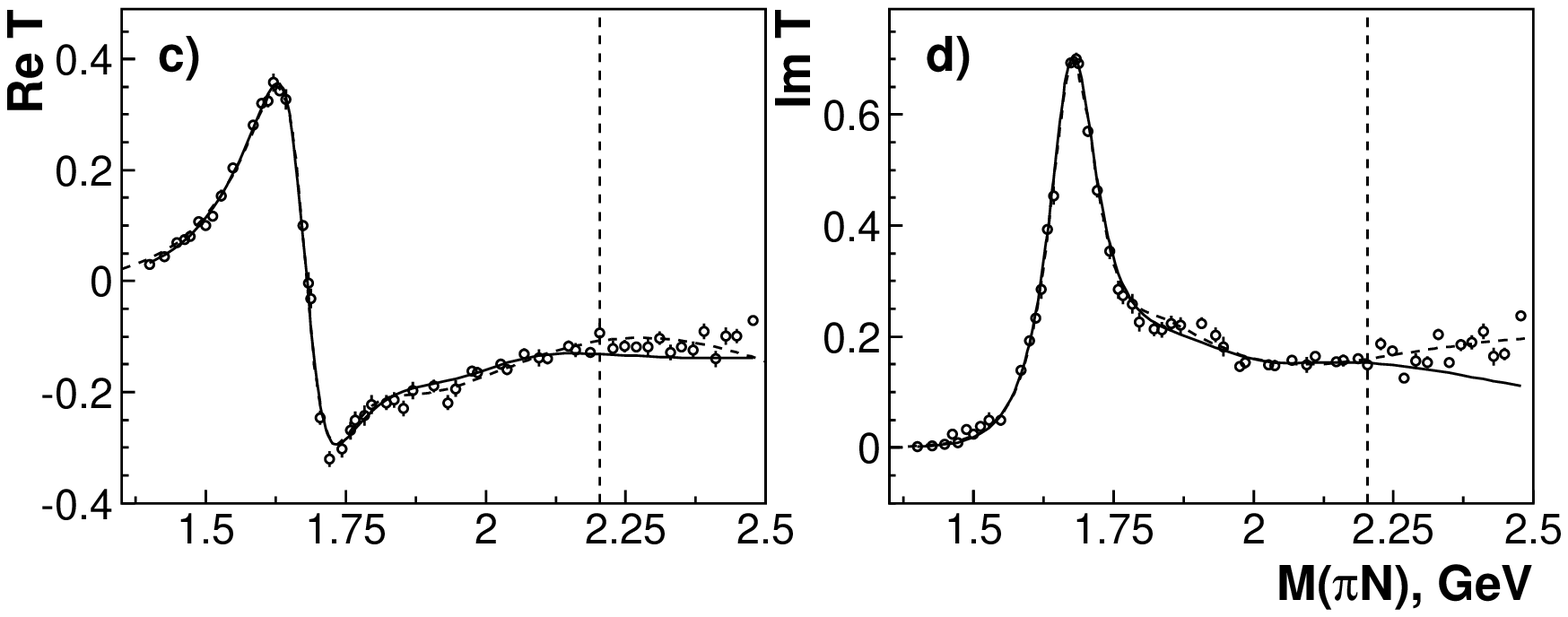,width=0.47\textwidth}
\epsfig{file=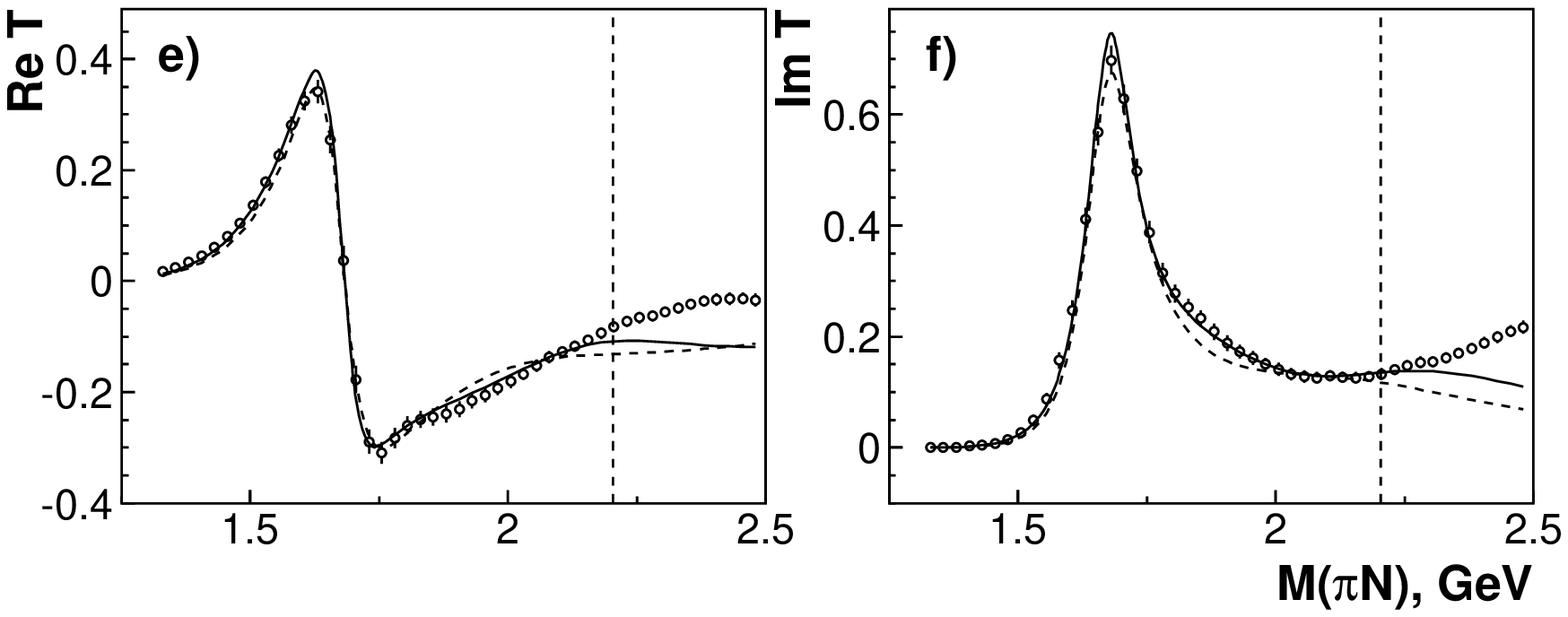,width=0.47\textwidth}
\caption{\label{pwa_f15}Real and imaginary part of the
$\frac12(\frac52) ^+$ partial wave amplitude. Points with error bars
are from SAID energy independent solution \cite{Arndt:2006bf} (a-d)
and from KH84 \cite{Hohler:1984ux} (f,e). The KH84 errors are
assumed to be $\pm 5$\%. (a,b) SAID energy dependent fit (dashed
curves) and fit BG2011-02 with two poles (solid curves). (c,d) SAID
fit (dashed curves) and fit BG2011-02 with three poles (solid
curves) (e,f) Fit BG2011-02 with two poles (dashed curves) and three
poles (solid curves). Points in the high-mass region, defined by the
vertical dashed line, were not included in the fit. \vspace{2mm}}
\centerline{\epsfig{file=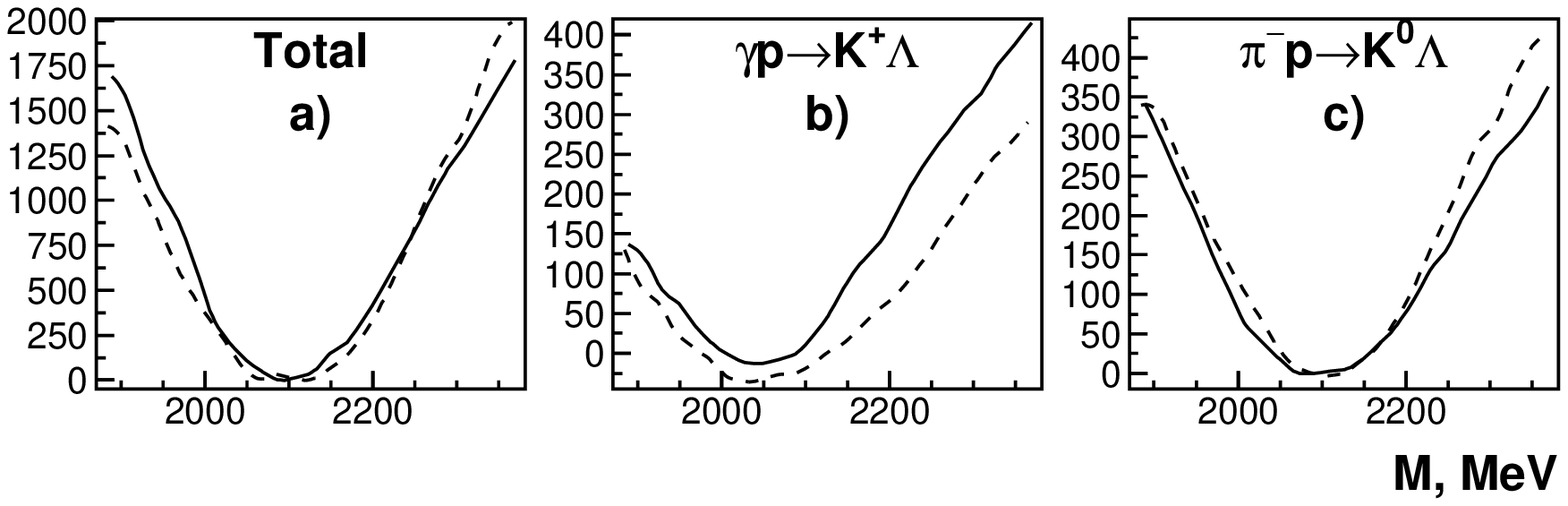,width=0.48\textwidth}}
\caption{\label{f15_ms}Mass scan in the $F_{15}$ wave above
$N(1680)F_{15}$: a) the change of total $\chi^2$ , b) the change of
$\chi^2$ value for the description of the $\gamma p\to K\Lambda$
data, differential cross section and recoil polarization, and c) the
change of $\chi^2$ for the description of the $\pi^- p\to K\Lambda$
data. The solution BG2011-02 is shown with solid curves and the
solution BG2011-01 with dashed curves.}
\end{figure}

The lowest $I(J^P)=\frac12(\frac52 ^+)$  state, $N(1680)F_{15}$, is
well established and its properties, including photoproduction
couplings, are known with a good precision. Real and imaginary part
of the amplitude derived from energy independent analysis
\cite{Arndt:2006bf} are shown in Fig.~\ref{pwa_f15}a,b. The
amplitude exhibits a beautiful resonance behavior due to
$N(1680)F_{15}$. We reproduce the resonance with properties fully
consistent with those given in \cite{Nakamura:2010zzi}.

The situation at higher masses is, however, far from being
satisfactory. A rather narrow state, with the mass $1882\pm 10$ and
width $95\pm 20$, was observed in the KH84 analysis of the elastic
$\pi N$ data \cite{Hohler:1984ux}. This state was confirmed by SAID
\cite{Arndt:2006bf}, although with a notably lower mass (1818 MeV).
In contrary, an observation of a rather broad state ($M\!=\!1903\pm
87$, $\Gamma\!=\!490\pm 310$ MeV) was reported from the combined
analysis of the $\pi N$ elastic data and $\pi N\to 2\pi N$ data
\cite{Manley:1992yb}. The state was not observed in the CM analysis
\cite{Cutkosky:1980rh}.

In \cite{Hohler:1984ux,Arndt:2006bf}, the evidence for this second
$F_{15}$ state is derived from the small structure in the amplitude
(Fig.~\ref{pwa_f15}) at about 1.9\,GeV which is present in the
amplitudes from KH and GWU. Including a second resonance in the
$F_{15}$ wave, the KH points were described with a good $\chi^2$,
the GWU points with a fair $\chi^2$.

Our coupled-channel analysis demands a second $5/2^+$ state in the
region above 1800 MeV even though at a considerably higher mass. In
a two-pole five-channel K-matrix parameterization of the $F_{15}$
partial wave, the position of the second pole was found to be at
$2050\pm 30-i235\pm 20$ (see Table~\ref{residues5p}). The elastic
$\pi N$ amplitude is described with a modest $\chi^2/N_{\rm
data}=4.72$ per data point (BG2011-02), see Fig.~\ref{pwa_f15}a,b.
The solution BG2011-01 produces a similar description, with
$\chi^2/N_{\rm data}=4.44$. Moreover we did not find a notable
difference for the parameterization of this wave in our solutions
BG2011-01 and BG2011-02.

The need for a second resonance above $N(1680)F_{15}$ can be
visualized in a mass scan. Mass scans with two K-matrix poles are
difficult to interpret, hence the $F_{15}$ partial wave is
parameterized as a sum of a one-pole K-matrix and a Breit-Wigner
amplitude. With this parameterization we performed a mass scan where
the mass of the Breit-Wigner amplitude was varied in defined steps,
while all other parameters of the fit were allowed to re-adjust in
each step. The  $\chi^2$ of the fit is thus a function of the
assumed Breit-Wigner mass. The need for a further $F_{15}$ resonance
above $N(1680)F_{15}$ can also be seen, e.g., from two fits - with
one and with two  $F_{15}$ resonances, respectively - shown in Fig.
\ref{pol_c}.

Fig.~\ref{f15_ms}a shows the $\chi^2$ as a function of the assumed
Breit-Wigner mass in the $I(J^P)=1/2(5/2 ^+)$ partial wave. A clear
minimum in $\chi^2$ is observed at about 2100\,MeV. Main $\chi^2$
changes stem from deteriorations of the fit to the differential
cross section and the recoil polarization for $\gamma p\to
K^+\Lambda$ and of $\pi^- p\to K^0\Lambda$ (Fig.~\ref{f15_ms}b,c).
In the former reaction, the $\chi^2$ changes in the double
polarization variables $C_x,C_z,O_x$, $O_z$ are small, below 20
units, and are not included in the plot. The minimum in all
distributions is well defined, although it is slightly lower in mass
for the photoproduction data than for pion induced reactions.

The structure in the $F_{15}$ amplitude at about 1.9\,GeV
(Fig.~\ref{pwa_f15} a,b) is not well described in the fit with two
K-matrix poles. Hence we introduced a three-pole five-channel
K-matrix to describe the $F_{15}$ partial wave and increased the
weight of the $F_{15}$ elastic amplitude points in the fit. The new
solutions reproduce the SAID $F_{15}$ amplitude with $\chi^2/N_{\rm
data}=1.82$ (BG2011-01) and $\chi^2/N_{\rm data}=1.91$ (BG2011-02):
see Fig.~\ref{pwa_f15}c,d. For the KH amplitude, the $\chi^2/N_{\rm
data}$ improved from 3.75 for the two-pole solution BG2011-02 to
0.73 for the three-pole solution BG2011-02): see
Fig.~\ref{pwa_f15}e,f. The total $\chi^2$ improved by 750 units,
mostly due to improvements in the description of elastic $F_{15}$
amplitude (320) and the $\gamma p\to \pi^0 p$ data (400). However,
we found two types of the three-pole solutions:

In the solution (a), the second pole is located in the mass region
1800-1950\,MeV and its imaginary part corresponds to a width of
$120-300$ MeV. This pole has not only a high inelasticity; more than
80\% of its decays must go into channels like $\rho N$ and $\omega
N$ since nearly no evidence is seen for this resonance in the
channels studied here (except $\gamma N,\pi N \to \pi N$) . The
third pole in this solution is shifted to a higher mass, by about 50
MeV. The third pole for this solution is given in
Table~\ref{residues5p} while the second pole is not, since its mass
is ill defined.

In solutions of type (b), the two highest poles come very close to
each other; both are located in the 1900-1980\,MeV region, and in
the complex plane one pole falls atop of the other one. The pole
position and couplings for these solutions are listed in
Table~\ref{residues5p}. If both poles correspond to real resonances,
one of the poles must have large couplings to $\rho N$ and $\omega N$
while the other couples significantly to $K\Lambda$.

In all solutions, using the SAID \cite{Arndt:2006bf} or KH84
\cite{Hohler:1984ux} $F_{15}$ amplitude, a two-pole or three-pole
K-matrix, one pole is required in the 1950-2100 mass region. It
leads to a highly significant $\chi^2$ improvement in reactions with
$K\Lambda$ in the final state. There is suggestive evidence for a
further pole between 1800 and 1950\,MeV: this state is mainly needed
to improve the description of the elastic amplitudes. Its existence
is supported by the analyses of H\"ohler \cite{Arndt:2006bf}, of
Manley {\it et. al} \cite{Manley:1992yb}, and, based on our
analysis, its existence seems rather likely. We do not claim that
this state must exist but we certainly cannot rule out the
possibility that it does exist. In the discussion, we refer to this
state as $N_{5/2^+}(1875)$.

\begin{table}[pt]
\caption{\label{residues5p}$F_{15}$ wave:  Pole positions and
residues of the transition amplitudes (given in MeV as $|r|/\Theta$
with $Res A=|r|\,e^{i\Theta}$). The phases are given in degrees. The
helicity couplings are given in 10$^{-3}$GeV$^{-1/2}$. The errors
are defined from the spread of results found in the respective
classes of solutions. The RPP values are given in parentheses. The
third pole in the three-pole solution (a) is ill defined. It is
located in the 1800-1950\,MeV mass range.}
\begin{footnotesize}
\renewcommand{\arraystretch}{1.10}
\bc
\begin{tabular}{lcc}
\hline\hline
\vspace{-0.3cm}\\
Solution           &\hspace{-2mm} 2 poles   & \hspace{-2mm} 3 poles (Sol. a)\\
\hline\hline
\vspace{-0.3cm}\\
State              &\hspace{-2mm}$N(2000)F_{15}$ & \hspace{-2mm}$N(2000)F_{15}$\\
\hline
\vspace{-0.3cm}\\
Re(pole)           &\hspace{-2mm}$2050\!\pm\!30$ ( - )    &  \hspace{-2mm}$2095^{+30}_{-60}$ ( - )                 \\
-2Im(pole)         &\hspace{-2mm}$470\!\pm\!40$  ( - )    &  \hspace{-2mm}$500\!\pm\!80$  ( - )                 \\
BW mass            &\hspace{-2mm}$2090\!\pm\!20$ ($\sim\!2000$)  & \hspace{-2mm}$2190\!\pm\!40$ ($\sim\!2000$)    \\
BW width           &\hspace{-2mm}$450\!\pm\!40$  ( - )    &  \hspace{-2mm}$550\!\pm\!100$  ( - )                 \\
$A(\pi N\to \pi N)$ &\hspace{-2mm}$20\!\pm\!4$\,/-$60\!\pm\!20^o$ & \hspace{-2mm}$33\!\pm\!10$\,/-$115\!\pm\!40^o$   \\
$A(\pi N \to K\Lambda)$   &\hspace{-2mm} ~$20\!\pm\!5$\,/$80\!\pm\!25^o$ & \hspace{-2mm} ~$16\!\pm\!8$\,/$70\!\pm\!30^o$  \\
$A(\pi N \to K\Sigma)$    &\hspace{-2mm} ~$15\!\pm\!5$\,/$70\!\pm\!25^o$   & \hspace{-2mm} ~$5\!\pm\!3$\,/ - \\
$A^{1/2}(\gamma p)$&\hspace{-2mm}$42\!\pm\!5$\,/-$30\!\pm\!10^o$& \hspace{-2mm}$35\!\pm\!20$\,/$20\!\pm\!20^o$ \\
$A^{3/2}(\gamma p)$&\hspace{-2mm}$48\!\pm\!5$\,/-$170\!\pm\!25^o$ & \hspace{-2mm}$50\!\pm\!20$\,/-$130\!\pm\!35^o$\\
\hline\hline
\vspace{-0.3cm}\\
Solution           &\hspace{-2mm} 3 poles (Sol. b)   &\hspace{-2mm} 3 poles (Sol. b)\\
\hline\hline
\vspace{-0.3cm}\\
State              &\hspace{-2mm}$N_{5/2}^+(1900)$ &\hspace{-2mm}$N(2000)F_{15}$ \\
\hline
\vspace{-0.3cm}\\
Re(pole)           &\hspace{-2mm}$1930\!\pm\!70$ ( - )    &\hspace{-2mm}$1920\!\pm\!70$                  \\
-2Im(pole)         &\hspace{-2mm}$400\!\pm\!40$  ( - )    &\hspace{-2mm}$540\!\pm\!40$                   \\
BW mass            &\hspace{-2mm}$2000\!\pm\!50$ ($\sim\!2000$)  &\hspace{-2mm}$2000\!\pm\!50$                  \\
BW width           &\hspace{-2mm}$410\!\pm\!40$  ( - )    &\hspace{-2mm}$500\!\pm\!40$                   \\
$A(\pi N\to\pi N)$ &\hspace{-2mm}$70\!\pm\!20$\,/-$100\!\pm\!80^o$ &\hspace{-2mm}$140\!\pm\!25$\,/-$80\!\pm\!40^o$   \\
$A^{1/2}(\gamma p)$&\hspace{-2mm}$50\!\pm\!15$\,/$150\!\pm\!40^o$&\hspace{-2mm}$45\!\pm\!10$\,/$55\!\pm\!40^o$ \\
$A^{3/2}(\gamma p)$&\hspace{-2mm}$65\!\pm\!10$\,/-$15\!\pm\!40^o$ &\hspace{-2mm}$75\!\pm\!15$\,/-$110\!\pm\!40^o$ \\
\hline\hline
\end{tabular}
\ec
\end{footnotesize}
\renewcommand{\arraystretch}{1.0}
\end{table}


In Fig.~\ref{sig_tot}e,f we show the contribution of the
$\frac12(\frac52 ^+)$ partial wave to the $\gamma p\to K^+\Lambda$
total cross section. The contribution is calculated starting from
the three-pole solution (a) and using the SAID $F_{15}$ $\pi N$
amplitude \cite{Arndt:2006bf}.

In Table \ref{residues5p} we present details of the fits: pole
position, Breit-Wigner parameters, residue of the elastic pole and
helicity amplitudes. For this partial wave the solutions BG2011-01
and BG2011-02 produce very similar results; the small differences
are included in the given errors.

\boldmath\subsubsection{$I(J^P)=\frac12(\frac72^+)$}\unboldmath

\begin{figure}[h]
\epsfig{file=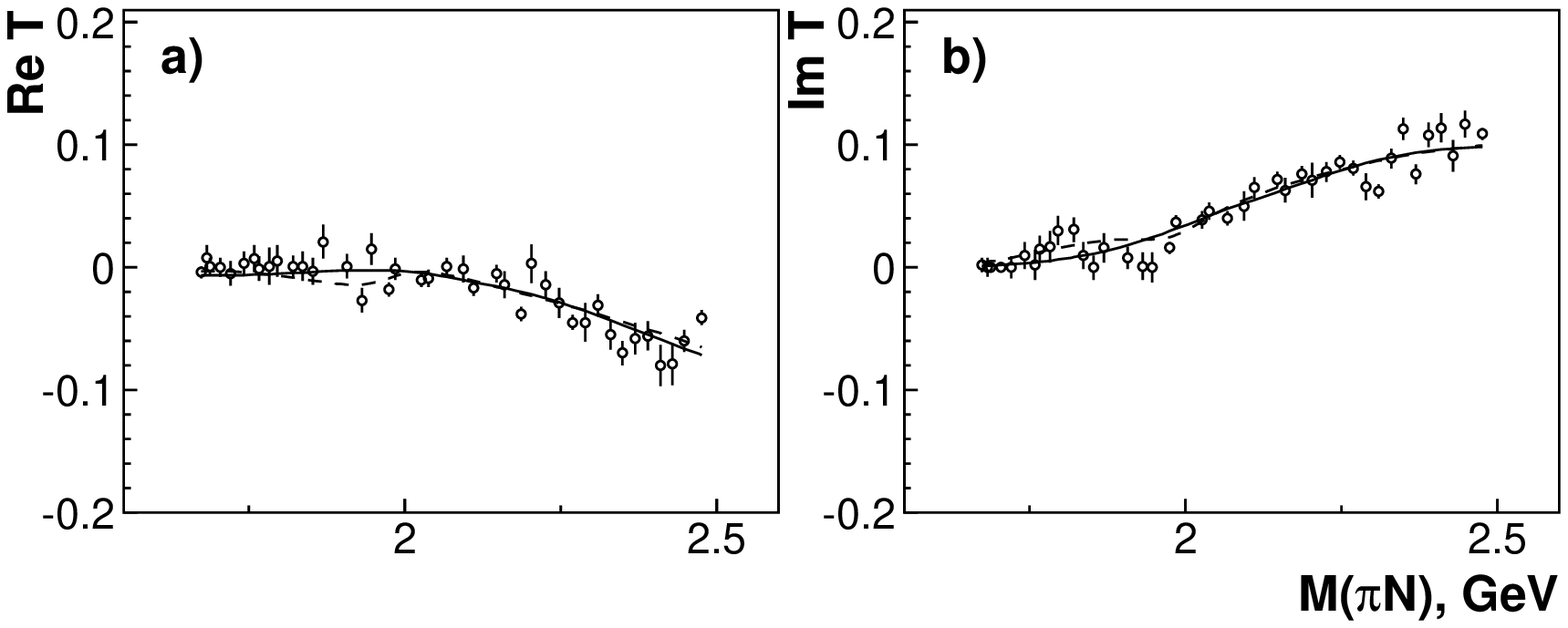,width=0.47\textwidth}
\epsfig{file=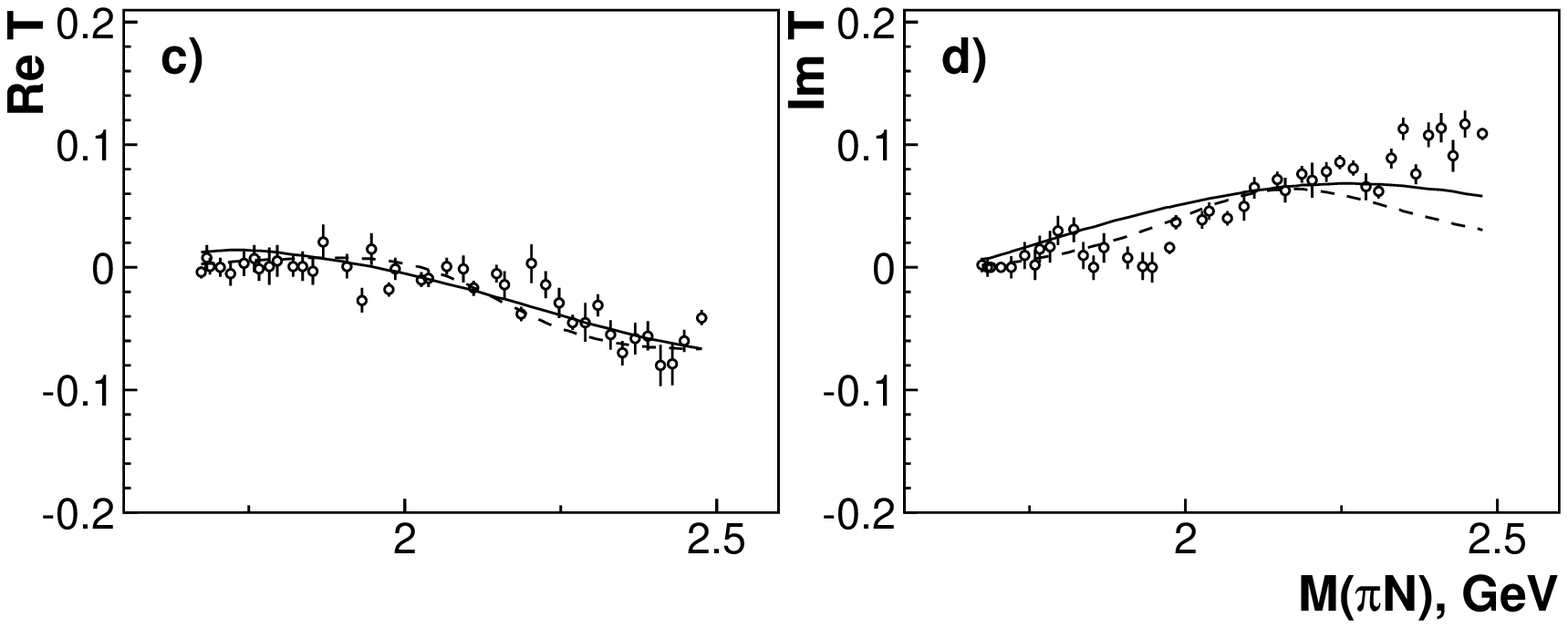,width=0.47\textwidth}
\caption{\label{pwa_f17}Real and imaginary part of the
$\frac12(\frac72)^+$ partial wave. Points with error bars are from
the SAID energy independent solution \cite{Arndt:2006bf}. The
solution BG2011-02 is shown as solid curve and BG2011-01 as dashed
curve (a, b). Same solutions without a high mass pole (around 2400
MeV) (c, d).}
\end{figure}

Observations of a $\frac12 (\frac72) ^+$ state in the 2\,GeV mass
region were reported from early KH84 \cite{Hohler:1984ux} and CM
\cite{Cutkosky:1980rh} analyses of the $\pi N$ elastic data, and
from an analysis which included data on $\pi N \to \pi \pi N$
\cite{Manley:1992yb}. Breit-Wigner mass and width were determined to
$M=2005\pm 150$\,MeV, $\Gamma=350\pm100$ in \cite{Hohler:1984ux}, to
$M=1970\pm 50$\,MeV, $\Gamma=350\pm120$ in \cite{Cutkosky:1980rh},
and to $M=2086\pm 28$\,MeV, $\Gamma=535\pm120$ in
\cite{Manley:1992yb}. The resonance is listed in RPP as
$N(1990)F_{17}$. The resonance was not seen in the analysis
presented in \cite{Arndt:2006bf} (the $I(J^P)=\frac12(\frac72) ^+$
amplitude is not given in the paper but on their web site). Indeed,
the $\frac12 (\frac72) ^+$ amplitude (see Fig.~\ref{pwa_f17}) shows
no significant structure.

\begin{figure}[h]
\centerline{\epsfig{file=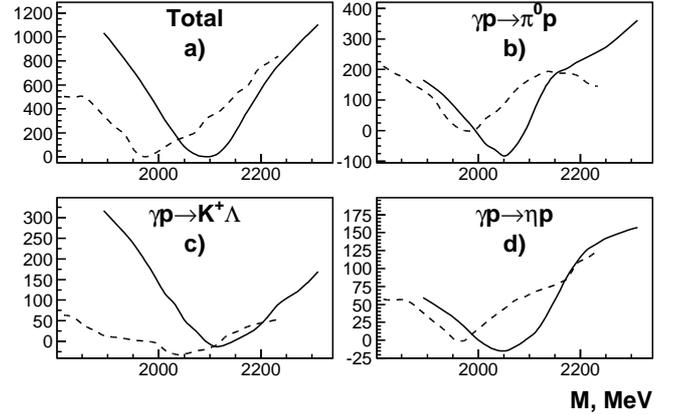,width=0.48\textwidth}}
\caption{\label{f17_scan}Mass scan in the $F_{17}$ wave with the
BG2011-02 (solid line) and BG2011-01 (dashed line) solutions. a)
Change of the total $\chi^2$ , b) $\chi^2$ change for $\gamma p\to
\pi^0 p$, c) for $\gamma p\to K^+\Lambda$, and d) for $\gamma p\to
\eta p$.}\vspace{-3mm}
\end{figure}

The results of the SAID and KH84 energy independent partial wave
analyses are shown in Fig.~\ref{pwa_f17}. A first attempt to
describe this amplitude jointly with our full data base with a
one-pole K-matrix parametrization failed. The main problem is that a
rather narrow state around 2\,GeV is required mainly by the data on
$\gamma p\to K\Lambda$ and $\gamma p\to \pi^0 p$, which is
incompatible with the $\pi N$ elastic amplitude \cite{Arndt:2006bf}
in this partial wave. The description of the elastic amplitude
improved considerably when a second K-matrix pole was included in
the fit. Its mass is not well defined, any pole in the mass between
2300 and 2500 MeV is suited (Fig~\ref{pwa_f17}a,b).

The main improvements due to a $F_{17}$ state are observed in the
description of the recoil polarization in $\gamma p\to K\Lambda$. In
this solution, the pole mass optimized for $1975\pm 15$\,MeV and the
width for $160\pm 30$\,MeV. In the solution BG2011-2, mass and width
were found to be $2100\pm 15$\,MeV and $260\pm 25$\,MeV. These two
solutions have very different helicity couplings for the $F_{17}$
state; therefore more precise polarization experiments should
distinguish between them. Further studies revealed that in both
classes of solutions, BG2011-01 and BG2011-02, a substitution of the
whole $F_{17}$ partial wave from one solution to the other one led
to a very similar fit quality. For simplicity, we associate the
low-mass resonance with BG2011-01 and the high-mass resonance with
BG2011-02.

\begin{table}[pt]
\caption{\label{residues7p}$F_{17}$ wave:  Pole positions and
residues of the transition amplitudes (given in MeV as $|r|/\Theta$
with $Res A=|r|\,e^{i\Theta}$). The phases are given in degrees. The
helicity couplings are given in 10$^{-3}$GeV$^{-1/2}$. The errors
are defined from the spread of results found in the respective
classes of solutions.}
\begin{footnotesize}
\renewcommand{\arraystretch}{1.10}
\bc
\begin{tabular}{lcc}
\hline\hline
\vspace{-0.3cm}\\
Solution           &\hspace{-2mm} BG2011-01             &\hspace{-2mm}BG2011-02          \\
\hline\hline
\vspace{-0.3cm}\\
State              &\hspace{-2mm}$N(1990)F_{17}$ &\hspace{-2mm}$N(1990)F_{17}$ \\
\hline
\vspace{-0.3cm}\\
Re(pole)           &\hspace{-2mm}$1975\!\pm\!15$   &\hspace{-2mm}$2100\!\pm\!15$                  \\
-2Im(pole)         &\hspace{-2mm}$160\!\pm\!30$    &\hspace{-2mm}$260\!\pm\!25$                   \\
BW mass            &\hspace{-2mm}$1990\!\pm\!15$   &\hspace{-2mm}$2105\!\pm\!15$                  \\
BW width           &\hspace{-2mm}$160\!\pm\!30$    &\hspace{-2mm}$260\!\pm\!25$                   \\
$A(\pi N\to\pi N)$ &\hspace{-2mm}$1.5\!\pm\!0.5$\,/$180\!\pm\!30^o$&\hspace{-2mm}$1\!\pm\!0.5$\,/$80\!\pm\!20^o$   \\
$A(\pi N \to K\Lambda)$   &\hspace{-2mm} ~$1\!\pm\!1$\,/$170\!\pm\!30^o$ & ~$0.5\!\pm\!0.5$\,/-$100\!\pm\!30^o$   \\
$A(\pi N \to K\Sigma)$    &\hspace{-2mm} ~$1\!\pm\!0.5$\,/$100\!\pm\!30^o$   & ~$0.3\!\pm\!0.3$\,/-$60\!\pm\!40^o$  \\
$A^{1/2}(\gamma p)$&\hspace{-2mm}$20\!\pm\!10$\,/$0\!\pm\!30^o$ &\hspace{-2mm}$55\!\pm\!15$\,/-$50\!\pm\!20^o$ \\
$A^{3/2}(\gamma p)$&\hspace{-2mm}$35\!\pm\!10$\,/$0\!\pm\!20^o$ &\hspace{-2mm}$55\!\pm\!20$\,/-$50\!\pm\!25^o$ \\
\hline\hline
\end{tabular}
\ec
\end{footnotesize}
\renewcommand{\arraystretch}{1.0}
\end{table}

A mass scan in the $F_{17}$ wave for the two solutions is shown in
Fig.~\ref{f17_scan}. In solution BG2011-02, all three reactions,
$\gamma p\to K\Lambda$, $\gamma p\to \pi^0 p$, and $\gamma p\to \eta
p$ contribute to the signal; in solution BG2011-01, the mass scan
shows significant minima for $p\pi^0$, $n\pi^+$, $n\eta $, and a
small dip only for $K^+\Lambda$.

The overall description of the data and the fit quality of the
elastic amplitudes hardly differ for our two solutions BG2011-01 and
BG2011-02; but the $F_{17}$ amplitudes have their lowest-mass pole
at a rather different position, at $(1975 - i\,80)$ or
$(2100-i\,130)$\,MeV, respectively.

\boldmath\subsubsection{$I(J^P)=\frac12(\frac92^+)$}\unboldmath

\begin{figure}[h]
\epsfig{file=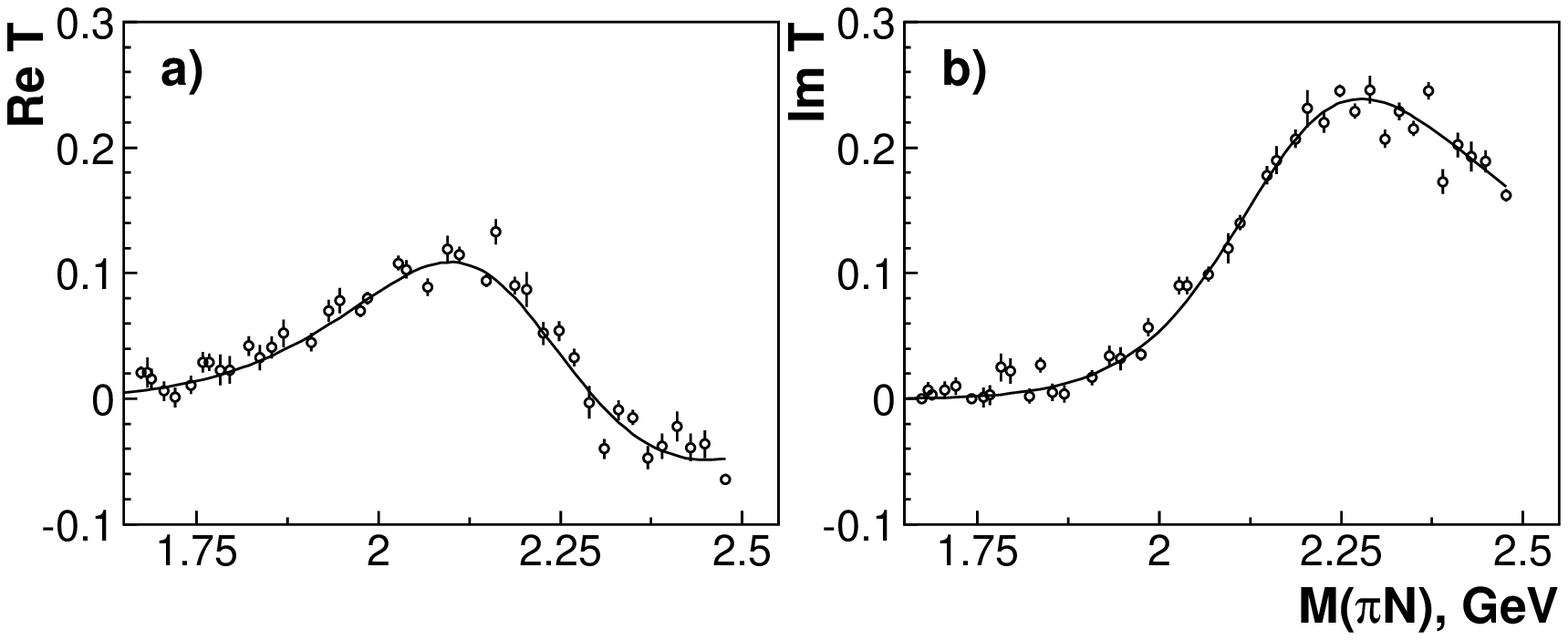,width=0.47\textwidth}
\epsfig{file=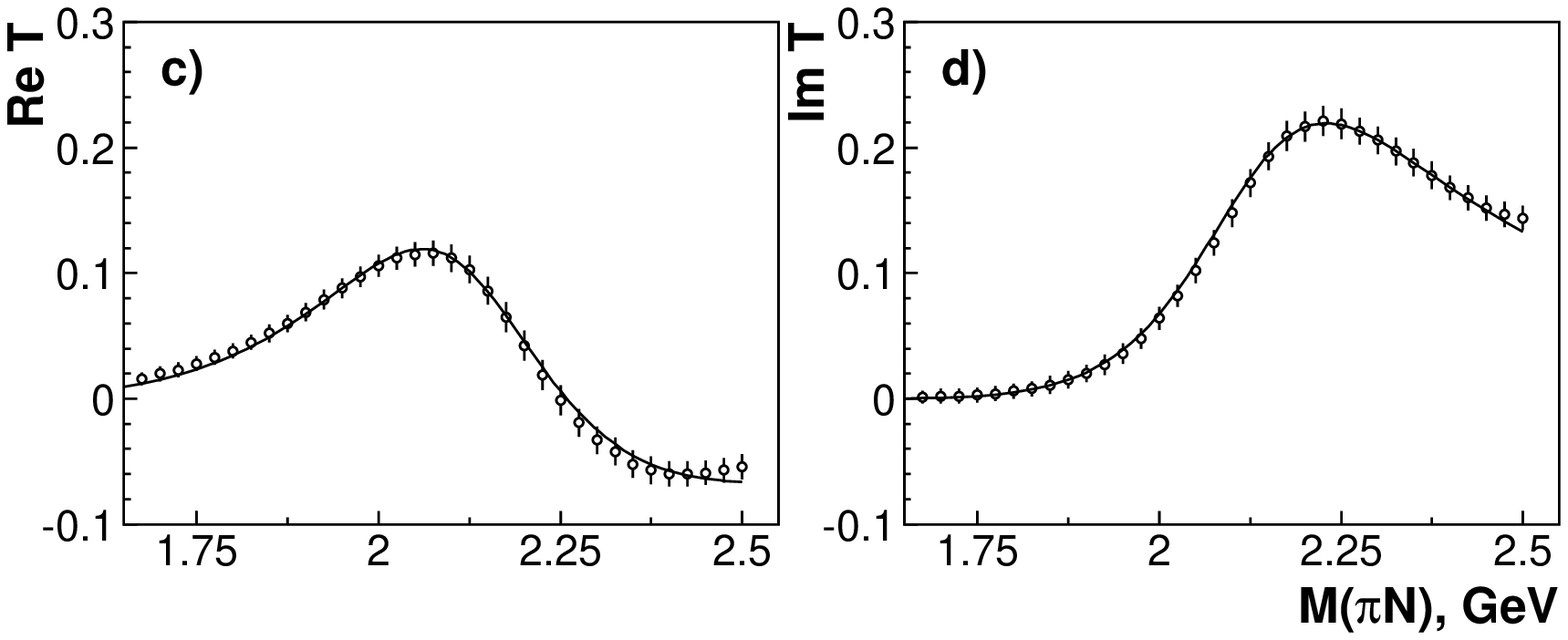,width=0.47\textwidth}
\caption{\label{pwa_h19}Real and imaginary part of the
$\frac12(\frac92) ^+$ partial wave. Points with error bars are from
the SAID energy independent solution \cite{Arndt:2006bf} (a,b) and
from KH84 \cite{Hohler:1984ux} (c,d). For KH84, $\pm 5$\% errors
were assumed.}
\end{figure}

The state $N(2220)H_{19}$ was clearly seen in the analysis of the
elastic data. We fitted these data as one-pole three-channel ($\pi
N$, $\pi\Delta$ and $\rho N$) K-matrix and included a $\pi N\to \pi
N$ non-resonant term. In the fits to the SAID \cite{Arndt:2006bf}
points, the non-resonant term can be taken as a constant while the
fit to the KH84 \cite{Hohler:1984ux} data, a more complicated form
(see section \ref{method}) was chosen. The description of the SAID
and KA84 results is shown in Fig.~\ref{pwa_h19}. We did not find any
appreciable contribution from the $N(2220)H_{19}$ to the
photoproduction data. The helicity couplings were optimized at very
small values and the quality of the description did not improved
after including this resonance in the photoproduction data. The pole
position and elastic residue for this state (averaged over two
solutions BG2011-01 and BG2011-02) is given in
Table~\ref{residues1}.

\begin{table}[pt]
\caption{\label{residues1}$H_{19}$ wave: Pole positions and residues
of the transition amplitudes (given in MeV as $|r|/\Theta$ with $Res
A=|r|\,e^{i\Theta}$). The phases are given in degrees. The helicity
couplings are given in $10^{-3}GeV^{-1/2}$. The errors are defined
from the spread of results found in both classes of solutions. The
RPP values are given in parentheses.}
\begin{footnotesize}
\renewcommand{\arraystretch}{1.1}
\bc
\begin{tabular}{lccc}
\hline \hline
\vspace{-0.3cm}\\
State              &\hspace{-2mm} &\multicolumn{2}{c}{$N(2220)H_{19}$}  \\
\hline
\vspace{-0.3cm}\\
Re(pole)           &\hspace{-2mm} &$2150\!\pm\!35$ &($\sim\!2170$) \\
-2Im(pole)         &\hspace{-2mm} &$440\!\pm\!40$ &($\sim\!480$)  \\
BW mass            &\hspace{-2mm} &$2200\!\pm\!50$&($\sim\!2250$) \\
BW width           &\hspace{-2mm} &$480\!\pm\!60$ &($\sim\!400$)  \\
$A(\pi N\to\pi N)$ &\hspace{-2mm} &$60\!\pm\!12$\,/-$58\!\pm\!12^o$& \\
$A^{1/2}(\gamma p)$&\hspace{-2mm}  &$<10$  &\\
$A^{3/2}(\gamma p)$&\hspace{-2mm}   &$<10$ &\\
\hline\hline
\end{tabular}
\ec
\end{footnotesize}
\renewcommand{\arraystretch}{1.0}
\end{table}

\subsection{Negative parity nucleon resonances}

The mass spectrum of negative-parity nucleon resonances starts with
the well known spin doublet $N(1535)S_{11}$, $N(1520)D_{13}$ in the
second, and the spin triplet $N(1650)S_{11}$, $N(1700)D_{13}$,
$N(1675)D_{15}$ in the third resonance region. Within SU(6), these
states belong to a 70-plet. The next candidates listed in
\cite{Nakamura:2010zzi} are above 2\,GeV and have only a RPP star
rating of one or two stars. When we included the new CBELSA-TAPS
data on $\gamma p\to K^0\Sigma^+$, our fit failed to describe them
with reasonable accuracy. We had to introduce two new negative
parity resonances below 2\,GeV (even though some evidence for both
these resonances has been reported before). These two resonances
also improved considerably our description of the new CLAS data on
$\gamma p\to K^+\Lambda$ \cite{McCracken:2009ra}. We call these two
states $N_{1/2^-}(1895)$ and $N_{3/2^-}$ $(1875)$. A search for a
resonance in the $(I)J^P=\frac12(\frac52^-)$ wave finds evidence at
a considerably higher mass, at (2060 MeV). Our fits confirm the
well-known $N(2190)G_{17}$ and $N(2250)G_{19}$ in the
$(I)J^P=\frac12(\frac72^-)$ and $\frac12(\frac92^-)$ waves.

\boldmath\subsubsection{$I(J^P)=\frac12(\frac12^-)$}\unboldmath

\begin{figure}[pb]
\centerline{\epsfig{file=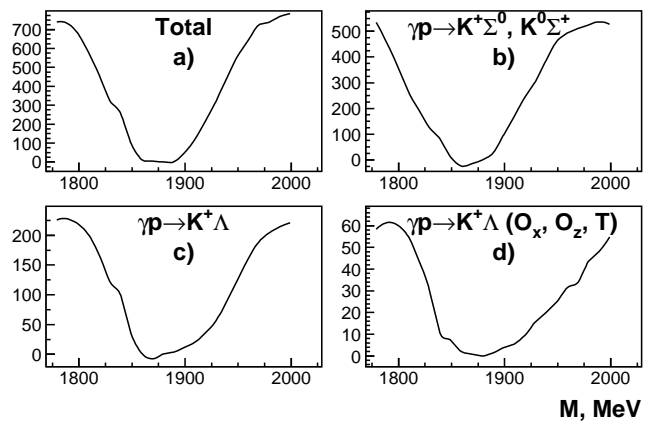,width=0.48\textwidth}}
\caption{\label{s11_scan}Mass scan of a third $S_{11}$ state. a) The
change of total $\chi^2$ , b) the change of $\chi^2$  value for the
description of the $\gamma p\to K^+\Sigma^0, K^0\Sigma^+$
differential cross section, c) the change of $\chi^2$ for the
description of the $\gamma p\to  K^+\Lambda$ differential cross
section and d) change of $\chi^2$ for the description of $O_x$,
$O_z$ and $T$ observables in the $\gamma p\to K\Lambda$ reaction.}
\end{figure}

The first evidence for a rather narrow $S_{11}$ state with the mass
$1880\pm 20$ and width $95\pm 30$ MeV was reported by KH84
\cite{Hohler:1984ux}. CM \cite{Cutkosky:1980rh} observed a
resonance, too, in this partial wave but at very different position:
at $M=2180\pm80$, $\Gamma=350\pm100$\,MeV. Manley {\it et al.}
reported a low-mass state, but now with a broad width
\cite{Manley:1992yb}. They reported $M=1928\pm 59$, $\Gamma=414\pm
157$\,MeV. In the later analyses of the GWU group
\cite{Arndt:2006bf}, there was no evidence at all for a third
$S_{11}$ state while Chen {\it et al.} \cite{Chen:2007cy} found
positive evidence at 1878\,MeV. These discrepant results are
summarizes by RPP under the one-star $N(2090)S_{11}$.

In the $S_{11}$ partial wave, our fits required, in addition to the
two-pole K-matrix parametrization of the low-energy part,
introduction of a further resonance at higher  mass. A Breit-Wigner
amplitude optimized at $M=1895\pm15$, $\Gamma=90^{+30}_{15}$ MeV, in
striking agreement with the result of the KH84 analysis.  The
resonance improved the description of differential cross section
\cite{McCracken:2009ra} and polarization observables $O_x,O_z$ and
$T$ \cite{Lleres:2007tx} from $\gamma p\to K^+\Lambda$. For the
latter data, a systematic improvement was obtained for two highest
energy bins at 1883 and 1908\,MeV. The mass scan of the $S_{11}$
state for the solution BG2011-02 is shown in Fig.~\ref{s11_scan}: it
demonstrates a clear minimum for photoproduction of $\Lambda$ and
$\Sigma$ hyperons, at the same mass and with a rather sharp minimum
as expected for a narrow state.

\begin{figure}[h]
\epsfig{file=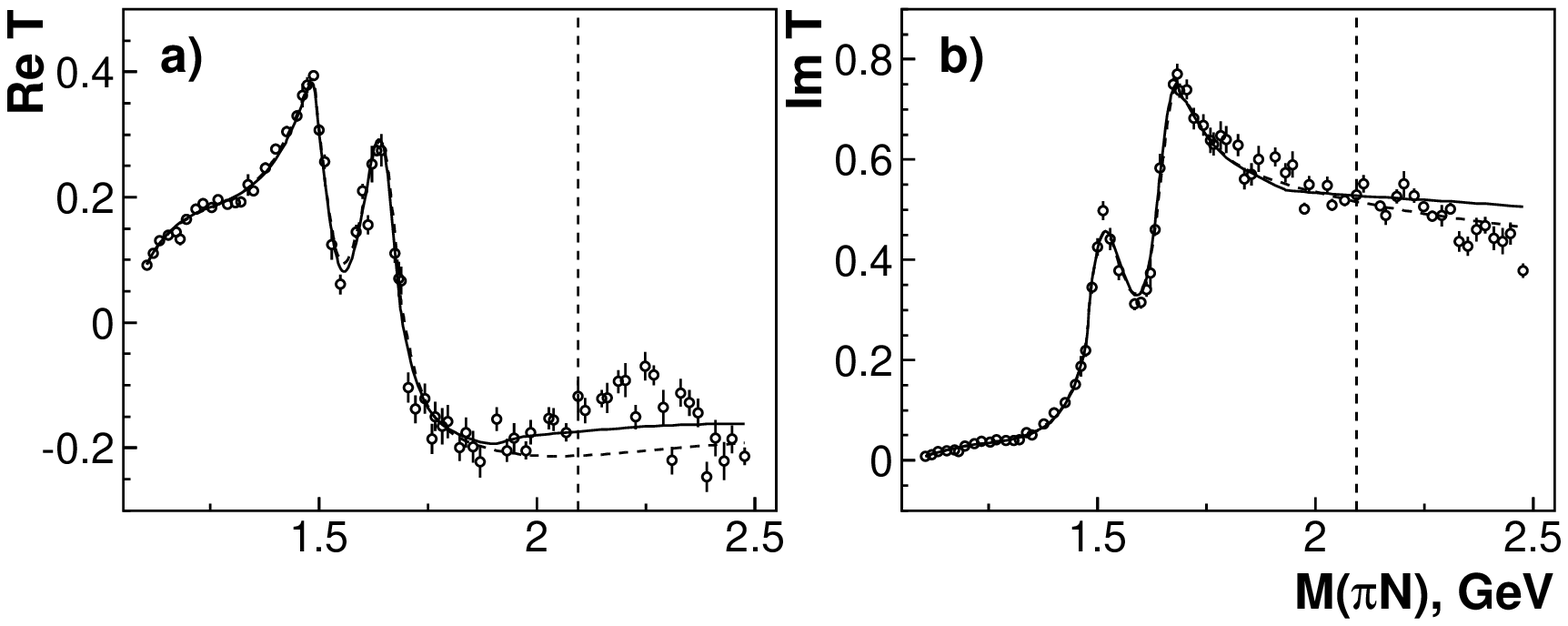,width=0.47\textwidth}
\epsfig{file=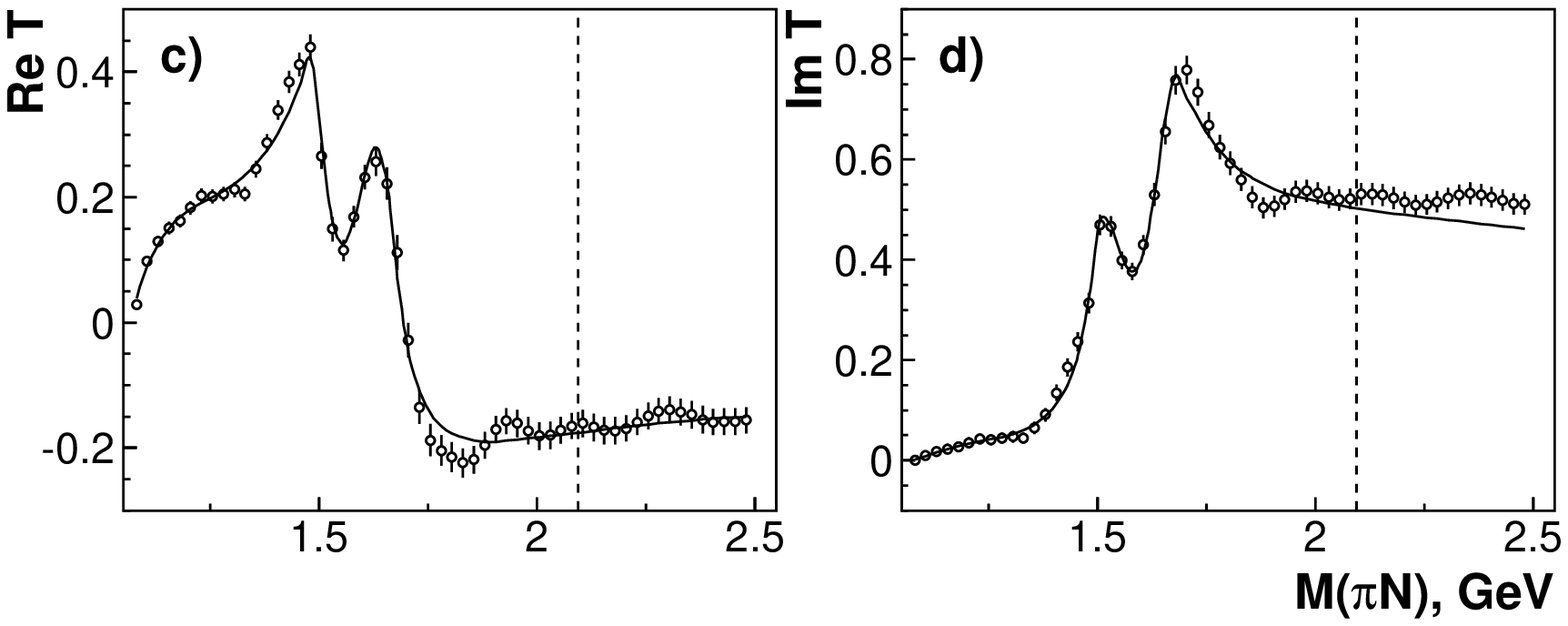,width=0.47\textwidth}
\caption{\label{pwa_s11}Real and imaginary part of the
$\frac12(\frac12) ^-$ partial wave. Points with error bars are from
the SAID energy independent solution \cite{Arndt:2006bf} (a,b) and
from KH84 \cite{Hohler:1984ux} (c,d). For KH84, $\pm 5$\% errors
were assumed. The SAID points are compared with our three pole
(solid lines) and two pole solutions (dashed lines). The KH84 points
are compared with three pole solution.  Points in the high-mass
region, defined by the vertical dashed line, were not included in
the fit.}
 \end{figure}

To check whether the existence of $N_{1/2-}(1880)$ is in conflict
with the elastic amplitudes extracted by the SAID and KA84 analyses
we included this resonance as third K-matrix pole and refitted our
full data base. The pole position of this state, found in the
three-pole six-channel K-matrix fit $\pi N$, $\eta N$, $K\Lambda$,
$K\Sigma$, $\pi\Delta$ and $\rho N$) as well as elastic residues and
photo-coupling are given in Table.~\ref{residuesneg}. The comparison
of our elastic $S_{11}$ amplitude with the SAID and KA84 energy
fixed analysis is shown in Fig.~\ref{pwa_s11}.

\boldmath
\subsubsection{$I(J^P)=\frac12(\frac32^-)$}
\unboldmath

A resonance structure at about 1900\,MeV in early SAPHIR data
\cite{Tran:1998qw} was explained \cite{Mart:1999ed} by including a
new resonance at 1895 MeV in the $I(J^P)=\frac12(\frac32^-)$ partial
wave. Our first analysis of data on photoproduction of $\pi^0$,
$\eta$, and of $\Lambda$ and $\Sigma$ hyperons
\cite{Anisovich:2005tf,Sarantsev:2005tg} suggested the existence of
two resonances with masses above 1800\,MeV in this wave which we
called $N_{3/2^-}(1875)$ and $N_{3/2^-}(2170)$. The existence of the
two resonances was required independently if the SAPHIR
\cite{Glander:2003jw} or CLAS \cite{Bradford:2005pt} data were used
in the fit. The Breit-Wigner parameters of $D_{13}(1870)$ were found
to be $M\!=\!1875\pm25$ MeV and $\Gamma\!=\!80\pm20$ MeV. The mass
scan for this state had a clear minimum in the 1870-1880\,MeV mass
region, for $\gamma p\to\pi N$, $K^+\Lambda$, and $K\Sigma$
\cite{Sarantsev:2005tg}.

\begin{figure}[pt]
\centerline{\epsfig{file=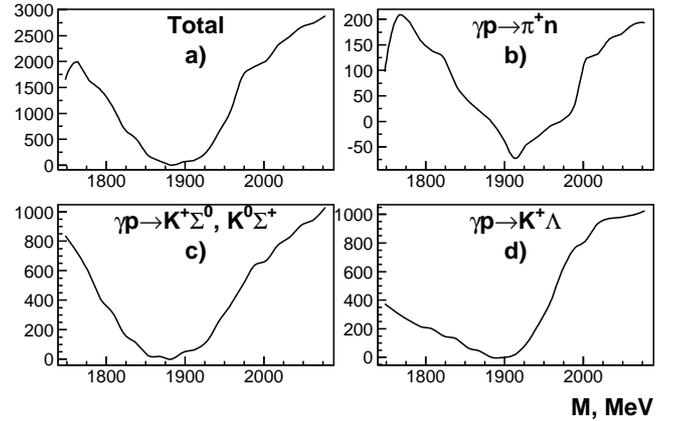,width=0.48\textwidth}}
\caption{\label{d13(1880)_scan}The mass scan of the $D_{13}(1870)$
state.  a) The change of total $\chi^2$ , b) the change of $\chi^2$
for the description of the $\gamma p\to \pi^+n$ reactions, c) the
change of $\chi^2$ for the description of the $\gamma p\to
K^+\Sigma^0,K^0\Sigma^+$ reactions and d) change of $\chi^2$ for the
description of the $\gamma p\to K^+\Lambda$ reactions.}
 \end{figure}

The present analysis uses a significantly richer data base, and the
parameters of these state are defined with a much better accuracy.
In a first parameterization, the $D_{13}$ partial wave was written
as an amplitude K-matrix with two poles and non-resonant (contact)
interactions; two Breit-Wigner amplitudes were added. The lower-mass
Breit-Wigner resonance was found at $1880\pm20$ MeV and width at
$200\pm25$ MeV. Its mass agrees precisely with our earlier finding,
however the new data demand a notably larger width. The mass of the
lower-mass Breit-Wigner amplitude was then scanned in a mass scan.
The mass scan is shown in Fig.~\ref{d13(1880)_scan}. It shows very
clear minima for these reactions due to contributions from the new
high statistic data
\cite{McCracken:2009ra,Ewald:2011a,Dugger:2009pn,Bradford:2005pt}.

\begin{figure}[pt]
\centerline{ \epsfig{file=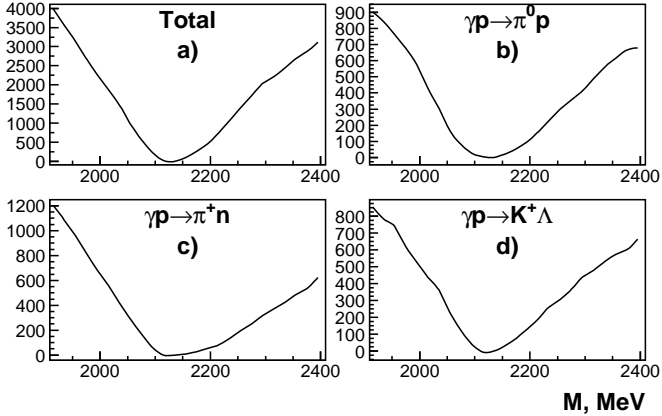,width=0.48\textwidth}}
\caption{\label{d13_scan}The mass scan of the fourth $D_{13}$ state.
a) The change of total $\chi^2$ , b) the change of $\chi^2$ for the
description of $\gamma p\to \pi^0p$, c) of $\gamma p\to \pi^+n$, and
d) of $\chi^2$ $\gamma p\to K^+\Lambda $. }
\end{figure}

Then, the $D_{13}$ partial wave was written as an amplitude with
three K-matrix poles, non-resonant (contact) interactions and one
Breit-Wigner amplitude. The latter amplitude corresponded to the
$D_{13}(2170)$ state. In \cite{Sarantsev:2005tg}, the high-mass
state $D_{13}(2170)$ was found at $M=2166^{+25}_{-50}$\,MeV and
$\Gamma=300\pm 65$\,MeV. In the present analysis this state, we find
its pole at $(2110\pm50) - i\,(170\pm23)$\,MeV. The mass scan for
this state is shown in Fig.~\ref{d13_scan}: the most significant
$\chi^2$ are seen in the photoproduction of the $N\pi$ and $K^+\Lambda$
final states.

\begin{figure}[h]
\epsfig{file=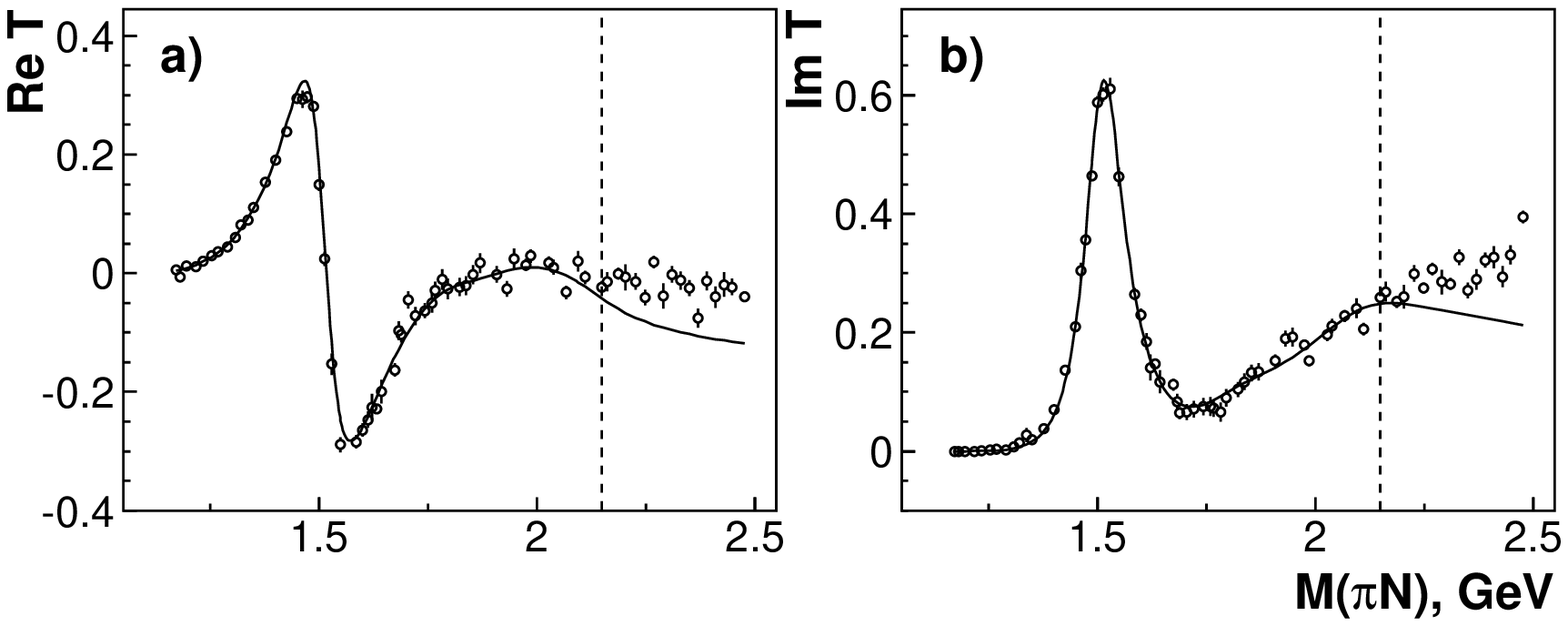,width=0.47\textwidth}
\epsfig{file=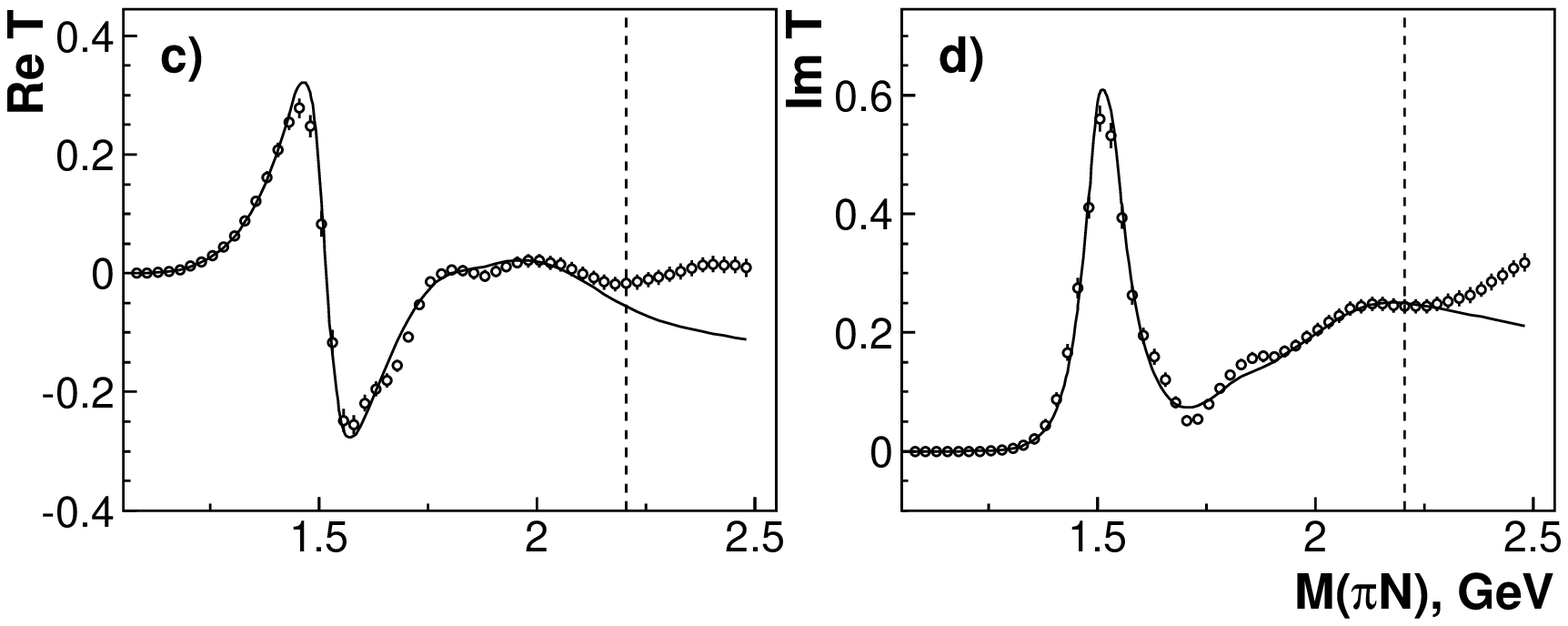,width=0.47\textwidth}
\caption{\label{pwa_d13}Real and imaginary part of the
$\frac12(\frac32) ^-$ partial wave.  Points with error bars are from
the SAID energy independent solution \cite{Arndt:2006bf} (a,b) and
from KH84 \cite{Hohler:1984ux} (c,d). For KH84, $\pm 5$\% errors
were assumed. Points in the high-mass region, defined by the
vertical dashed line, were not included in the fit. }
\end{figure}

Finally, a four-pole K-matrix was used. Pole positions, resi\-dues,
and helicity amplitudes for both resonances are given in Table
\ref{residuesneg}. A comparison of our elastic amplitude with the
SAID and KA84 amplitudes is shown in Fig.~\ref{pwa_d13}. It is seen
that in the fitted region our four-pole K-matrix describes these
elastic amplitudes rather well. However our analysis suggest the
possibility that at least one further $D_{13}$ resonance at a higher
mass might exist.

\begin{table}[pt]
\caption{\label{residuesneg}Negative parity waves: Pole positions
and residues of the transition amplitudes (given in MeV as
$|r|/\Theta$ with $Res A=|r|\,e^{i\Theta}$). The phases are given in
degrees. The helicity couplings are given in $10^{-3}GeV^{-1/2}$.
BG2011-01 and BG2011-02 yield consistent results, the errors are
hence defined from the spread of results found in the both classes
of solutions. The RPP values are given in parentheses.}
\begin{footnotesize}
\renewcommand{\arraystretch}{1.10}
\bc
\begin{tabular}{lcc}
\hline \hline
\vspace{-0.3cm}\\
State              &\hspace{-2mm} $N_{1/2^-}(1895)$                &$N_{3/2^-}(1875)$                 \\
\hline
\vspace{-0.3cm}\\
Re(pole)           &\hspace{-2mm}$1900\!\pm\!15$ ( - )  &$1860\!\pm\!25$ ( - )           \\
-2Im(pole)         &\hspace{-2mm}$90^{+30}_{-15}$ ( - )    &$200\!\pm\!20$  ( - )           \\
BW mass            &\hspace{-2mm}$1895\!\pm\!15$( - )   &$1880\!\pm\!20$ ( - )           \\
BW width           &\hspace{-2mm}$90^{+30}_{-15}$ ( - )    &$200\!\pm\!25$  ( - )           \\
$A(\pi N\to\pi N)$ &\hspace{-2mm}$0.5\!\pm\!0.5$\,/ -  &$ 2.5\!\pm\!1$\,/ -\\
$A(\pi N \to \eta N)$     &\hspace{-2mm} ~$2.5\!\pm\!1.5$\,/$40\!\pm\!20^o$  & \\
$A(\pi N \to K\Lambda)$   &\hspace{-2mm} ~$2\!\pm\!1$\,/-$90\!\pm\!30^o$ & ~$1.5\!\pm\!1$\,/ -   \\
$A(\pi N \to K\Sigma)$    &\hspace{-2mm} ~$3\!\pm\!2$\,/$40\!\pm\!30^o$   & ~$6\!\pm\!2$\,/$180\!\pm\!50^o$  \\
$A^{1/2}(\gamma p)$&\hspace{-2mm}$12\!\pm\!6$\,/$120\!\pm\!50^o$  &$ 18\!\pm\!8$\,/-$100\!\pm\!60^o$  \\
$A^{3/2}(\gamma p)$&\hspace{-2mm}~                          &$10\!\pm\!4$\,/$180\!\pm\!30^o$ \\
\hline
\vspace{-0.3cm}\\
State              &\hspace{-2mm}$N_{3/2^-}(2150)$                   & $N_{5/2^-}(2060)$  \\
\hline
\vspace{-0.3cm}\\
Re(pole)           &\hspace{-2mm}$2110\!\pm\!50$ ( - )                &$2040\!\pm\!15$ ( - ) \\
-2Im(pole)         &\hspace{-2mm}$340\!\pm\!45$  ( - )                &$390\!\pm\!25$ ( - )  \\
BW mass            &\hspace{-2mm}$2150\!\pm\!60$ ( - )                &$2060\!\pm\!15$( - ) \\
BW width           &\hspace{-2mm}$330\!\pm\!45$  ( - )                &$375\!\pm\!25$ ( - )  \\
$A(\pi N\to\pi N)$ &\hspace{-2mm}$13\!\pm\!3$\,/-$20\!\pm\!10^o$   &$19\!\pm\!5$\,/-$125\!\pm\!20^o$ \\
$A(\pi N \to K\Lambda)$   &\hspace{-2mm} ~$5\!\pm\!2$\,/$100\!\pm\!30^o$ & ~$1\!\pm\!0.5$\,/$20\!\pm\!60^o$   \\
$A(\pi N \to K\Sigma)$    &\hspace{-2mm} ~$3\!\pm\!2$\,/-$50\!\pm\!40^o$   & ~$7\!\pm\!4$\,/-$70\!\pm\!30^o$  \\
$A^{1/2}(\gamma p)$&\hspace{-2mm}$125\!\pm\!45$\,/-$55\!\pm\!20^o$     &$65\!\pm\!15$\,/$15\!\pm\!8^o$  \\
$A^{3/2}(\gamma p)$&\hspace{-2mm}$150\!\pm\!60$\,/-$35\!\pm\!15^o$ &$55^{+15}_{-35}$\,/$15\!\pm\!10^o$ \\
\hline
\vspace{-0.3cm}\\
State              &\hspace{-2mm} $N(2190)G_{17}$               & $N(2250)G_{19}$                 \\
\hline
\vspace{-0.3cm}\\
Re(pole)           &\hspace{-2mm}$2150\!\pm\!25$ ($\sim\!2075$) & $2195\!\pm\!45$ ($\sim\!2200$) \\
-2Im(pole)         &\hspace{-2mm}$330\!\pm\!30$ ($\sim\!450$)   & $470\!\pm\!50$ ($\sim\!450$)   \\
BW mass            &\hspace{-2mm}$2180\!\pm\!20$($\sim\!2190$)  & $2280\!\pm\!40$ ($\sim\!2275$) \\
BW width           &\hspace{-2mm}$335\!\pm\!40$ ($\sim\!500$)   & $520\!\pm\!50$ ($\sim\!500$)   \\
$A(\pi N\to\pi N)$ &\hspace{-2mm}$30\!\pm\!5$\,/$30\!\pm\!10^o$ & $26\!\pm\!5$\,/$-38\!\pm\!25^o$\\
$A(\pi N \to K\Lambda)$   &\hspace{-2mm} ~$4.5\!\pm\!2$\,/$20\!\pm\!15^o$ &    \\
$A(\pi N \to K\Sigma)$    &\hspace{-2mm} ~$7\!\pm\!3$\,/$90\!\pm\!30^o$   &   \\
$A^{1/2}(\gamma p)$&\hspace{-2mm}$63\!\pm\!7$\,/-$170\!\pm\!15^o$ & $<10$               \\
$A^{3/2}(\gamma p)$&\hspace{-2mm}$35\!\pm\!20$\,/$25\!\pm\!10^o$& $<10$               \\[0.2ex]
\hline\hline
\end{tabular}
\ec
\end{footnotesize}
\renewcommand{\arraystretch}{1.0}
\end{table}

\boldmath\subsubsection{$I(J^P)=\frac12(\frac52^-)$}\unboldmath

\begin{figure}[pt]
\epsfig{file=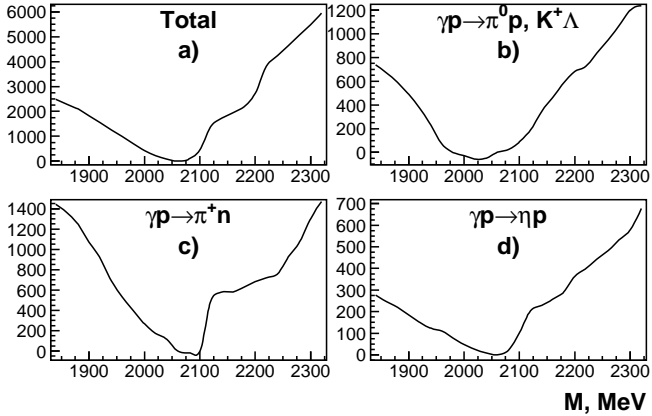,width=0.48\textwidth}
\caption{\label{d15_scan}The mass scan of the second $D_{15}$ state.
Changes of the $\chi^2$ value: a) for all fitted data, b) for the
$\gamma p\to \pi^0 p$  and $\gamma p\to K^+\Lambda$ data, c) for the
$\gamma p\to \pi^+ n$ differential cross section, d) for the $\gamma
p\to \eta p$ differential cross section. }
\end{figure}
The RPP \cite{Nakamura:2010zzi} lists one high-mass resonance in the
$I(J^P)=\frac12(\frac52^-)$ partial wave, the two-star
$N(2200)D_{15}$. It was reported by \cite{Hohler:1984ux} at $(M,
\Gamma)$=(2228$\pm$30, 310$\pm$50)\,MeV and by
\cite{Cutkosky:1980rh} at (2180$\pm$80, 400$\pm$100)\,MeV. In our
earlier analysis of the differential cross section on $\eta$
photoproduction \cite{Bartholomy:2004uz} we reported the observation
of a new $D_{15}$ resonance with a mass of about 2060\,MeV. The
state provided a dominant contribution to the $\gamma p\to \eta p$
cross section around 2050\,MeV; possibly these observations are
related. In the present solution, the contribution of this resonance
to the $\gamma p\to \eta p$ cross section remains practically
unchanged, but we find also a significant contribution from this
state to $\gamma p\to K^+\Lambda$. The new high statistic data on
the differential cross section and recoil asymmetry
\cite{McCracken:2009ra} are much better described when this
resonance is included. Also, the $\pi N$ coupling of this state
optimizes at a rather large value; it is of the same order of
magnitude as the $\eta N$ coupling constant. As a consequence, the
mass scan of the $\frac12(\frac52^-)$ wave demonstrates very clear
minima for the fit to $\gamma p\to \eta N$, to $\gamma p\to
K^+\Lambda$ and to both $\gamma p\to \pi N$ reactions (see
Fig.~\ref{d15_scan}).

\begin{figure}[h]
\epsfig{file=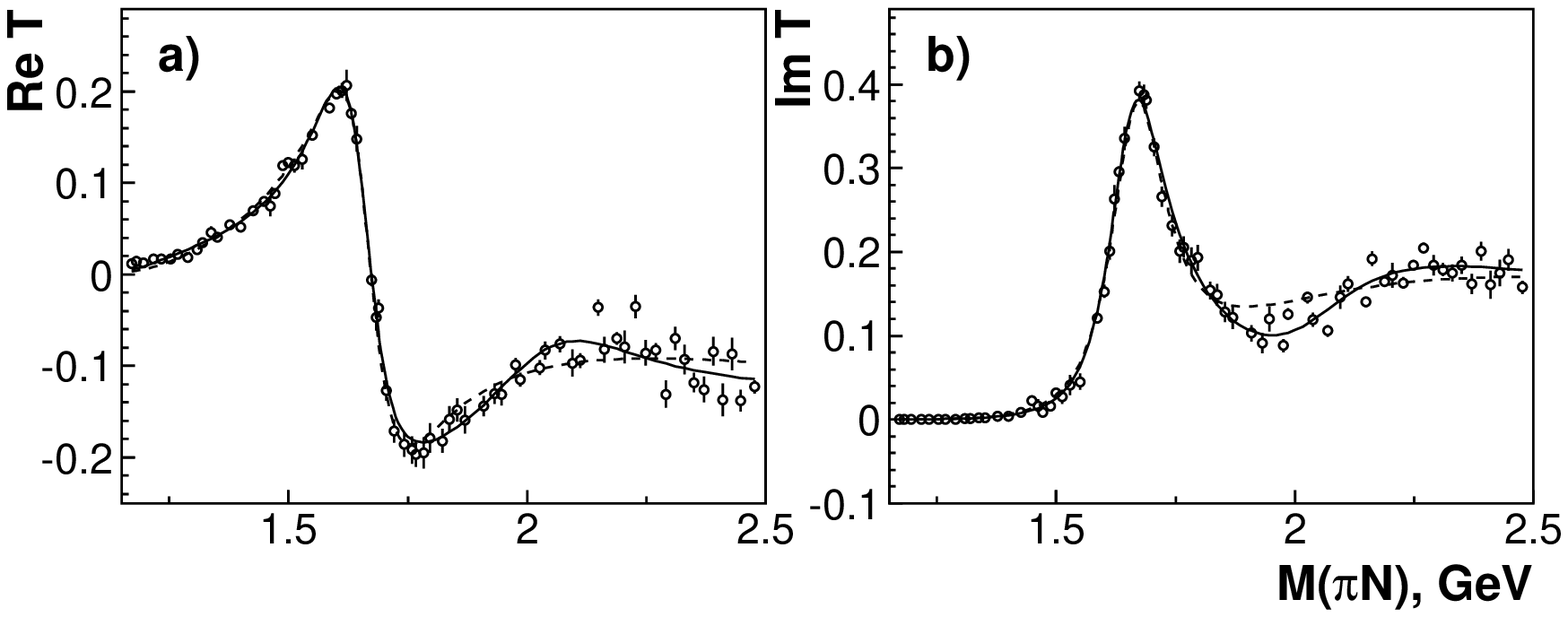,width=0.47\textwidth}
\epsfig{file=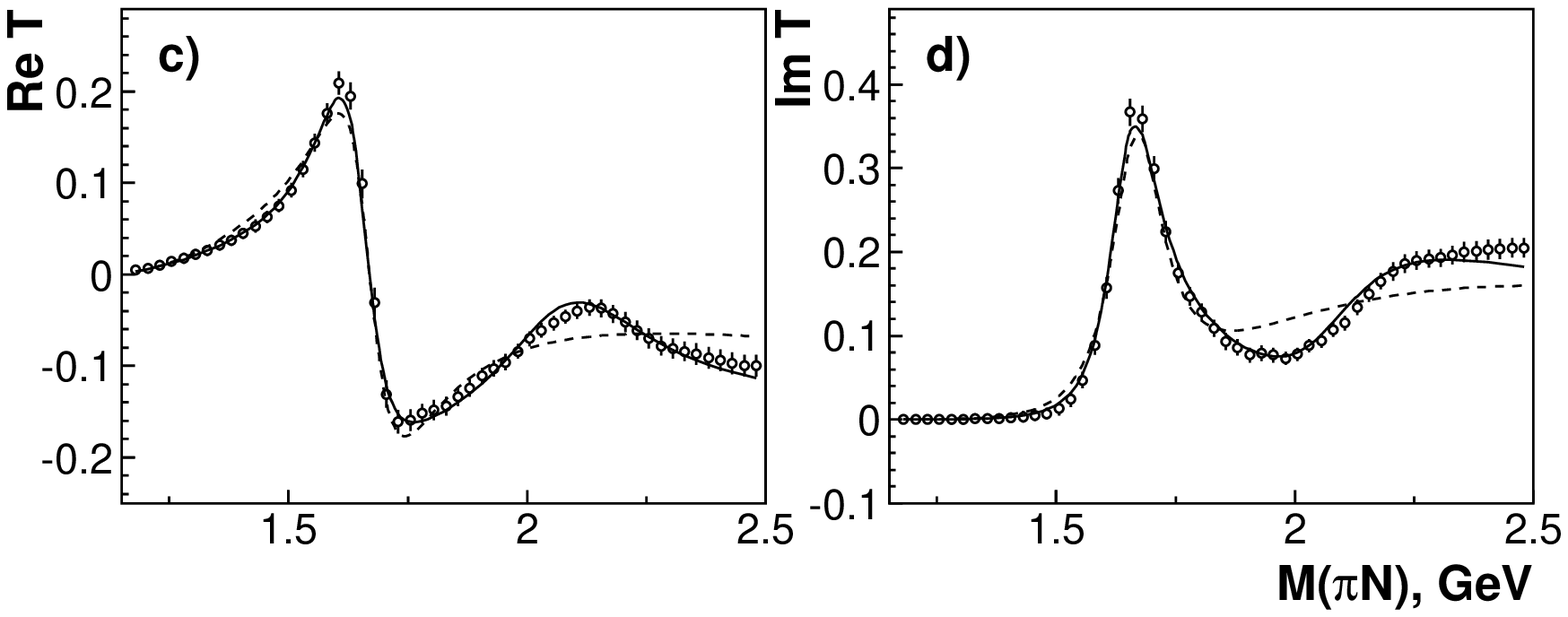,width=0.47\textwidth}
\caption{\label{pwa_d15}Real and imaginary part of the
$\frac12(\frac52) ^-$ partial wave.  Points with error bars are from
the SAID energy independent solution \cite{Arndt:2006bf} (a,b) and
from KH84 \cite{Hohler:1984ux} (c,d). For KH84, $\pm 5$\% errors
were assumed. The solid curves show the result of two pole K-matrix
parameterization and the dashed curves, the result of the one-pole
parameterization from BG2011-02.}
\end{figure}

The fit with the $N_{5/2-}(2060)$ state included as the second
K-matrix pole reproduces perfectly the $\pi N$ elastic amplitudes
extracted by the SAID and KA84 groups: see Fig.\ref{pwa_d15}. If the
second pole is excluded from the fit, the high energy region can not
be fitted well even after including energy dependent non-resonant
terms. The problem of such a parameterization are demonstrated in
Fig.~\ref{pwa_d15}. Although the elastic data indicative
the existence of a resonance above 2 GeV, the pole position of this
state is defined by the photoproduction data.

\boldmath\subsubsection{$I(J^P)=\frac12(\frac72^-)$}\unboldmath

\begin{figure}[h]
\epsfig{file=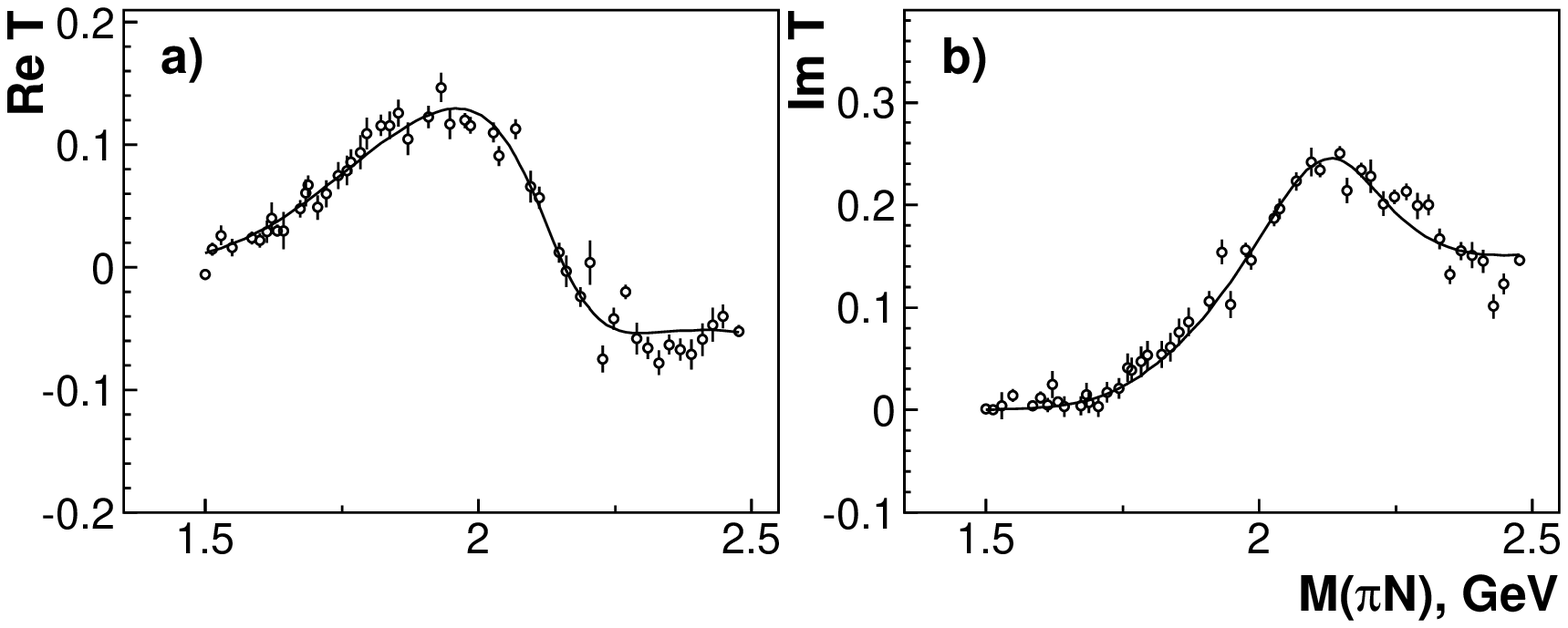,width=0.47\textwidth}
\epsfig{file=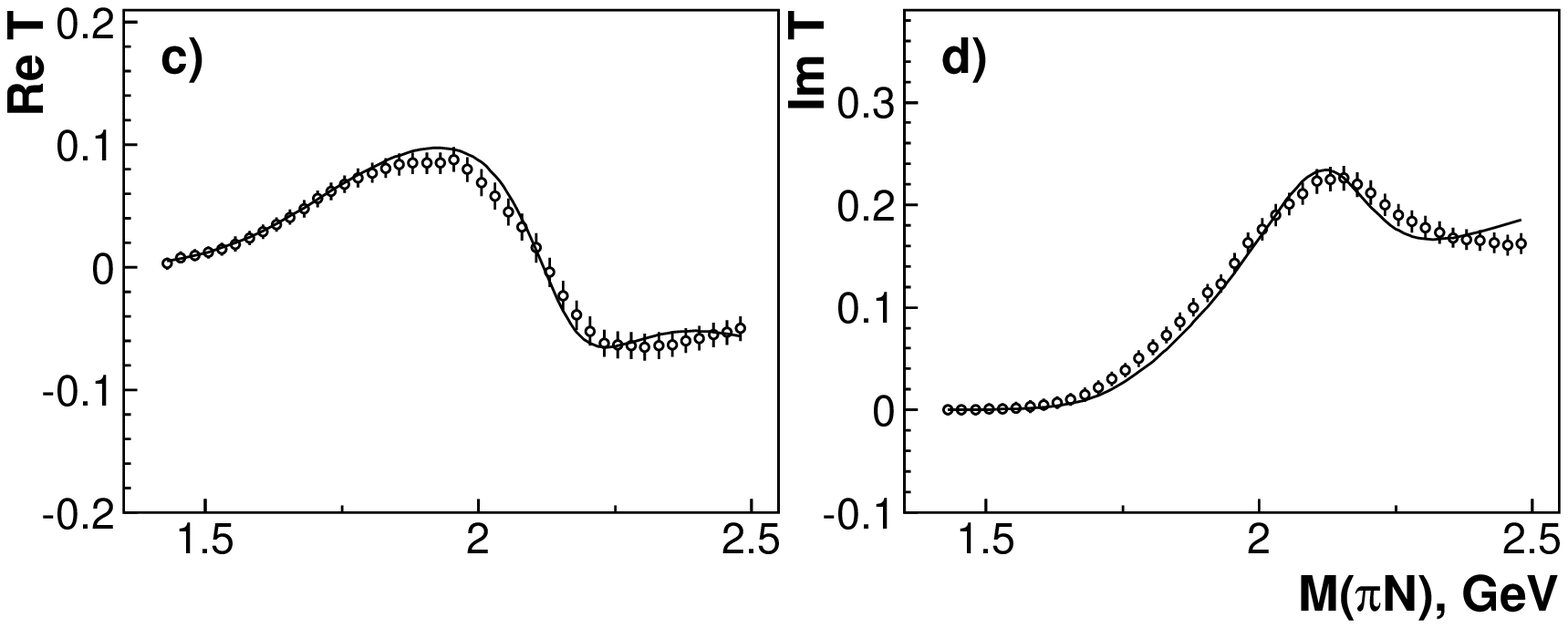,width=0.47\textwidth}
\caption{\label{pwa_g17}Comparison of the result of energy
independent analysis by SAID \cite{Arndt:2006bf} (a, b) and KH84
\cite{Hohler:1984ux} (c,d) for the $\frac12(\frac72) ^-$ partial
wave with the BG2011-02 solution. For KH84, $\pm 5$\% errors were
assumed.}
 \end{figure}

A resonance in the $I(J^P)=\frac12(\frac72^-)$ partial wave requires
internal orbital angular momenta of $L=3$; hence the lowest-mass
resonance can be expected to have a mass of at least 2\,GeV. Indeed,
a resonance in this partial wave is clearly seen from the behavior
of the elastic $\pi N$ scattering amplitude. The amplitudes of the
KH84 \cite{Hohler:1984ux} and analyses \cite{Arndt:2006bf} are shown
in Fig.~\ref{pwa_g17} as points with error bars. First, the real
part crosses zero and the imaginary part reaches its maximum
slightly above 2100 MeV, and second, the amplitude real part
increases slowly starting from 1500 MeV indicating a broad
structure. As the result, a one pole parameterization of the elastic
amplitude \cite{Arndt:2006bf} found a rather broad resonance with
the mass $2152.4\pm 1.4$ MeV and width $484\pm 13$ MeV. Other
analyses also found a resonance in this region but with masses which
cover the region 2100-2200 MeV and widthes between 300 and 700 MeV
(see references in RPP \cite{Nakamura:2010zzi}).

An introduction of a $\frac12(\frac72) ^-$ state in the mass region
2100 MeV as a Breit-Wigner amplitude significantly improves the
description of the $\gamma N\to K\Lambda$ differential cross section
and recoil asymmetry \cite{McCracken:2009ra}. Contributions of this
partial wave to the $\gamma p\to K\Lambda$ total cross section are
shown in Fig.~\ref{sig_tot}. This state also contributed notably to
the pion photoproduction. In our fits, the mass of the state was
found to be compatible with the RPP numbers \cite{Nakamura:2010zzi},
the width was optimized slightly above 300\,MeV. The fits required
some contribution from non-resonant transitions (parameterized as constants)
from $\gamma p\to \pi\Delta(1232)$ and $\rho N$ with the lowest
orbital momentum.

The comparison of our $\pi N$
elastic amplitude with the energy independent solution
\cite{Arndt:2006bf} is shown in Fig.~\ref{pwa_g17}. Our amplitude
has a pole at $2150-i\,165$ MeV. This pole provides a good description
of the $K\Lambda$ and $\pi N$ photoproduction data and is compatible
with the energy independent solution for the elastic amplitude
\cite{Arndt:2006bf} up to 2.4\,GeV.

\begin{figure}[h]
\centerline{\epsfig{file=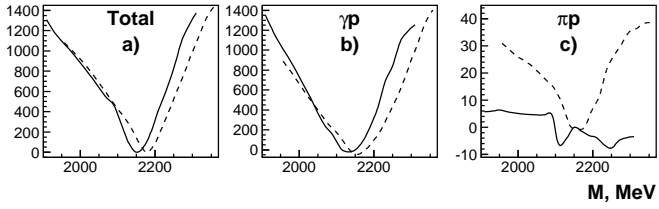,width=0.48\textwidth}}
\caption{\label{g17_scan}Mass scan in the $\frac12(\frac72) ^-$ wave
with the solution BG2011-02 (solid curves) and with the
solution BG2011-01 (dashed curves). a) Change of the total $\chi^2$
, b) change of the $\chi^2$ value for all photoproduction reactions and
c) change of the $\chi^2$ value for all pion induced inelastic
reactions.}
\end{figure}

A mass scan in the $\frac12(\frac72) ^-$ wave shows very good minima
for the total log likelihood value and for the sum of
photoproduction contributions (see Fig.~\ref{g17_scan}) while this
state improves the description of the pion-induced inelastic
reactions only marginally; the mass scan shows a small minimum for
the solution BG2011-01 only.

\boldmath\subsubsection{$I(J^P)=\frac12(\frac92^-)$}\unboldmath

\begin{figure}[pt]
\epsfig{file=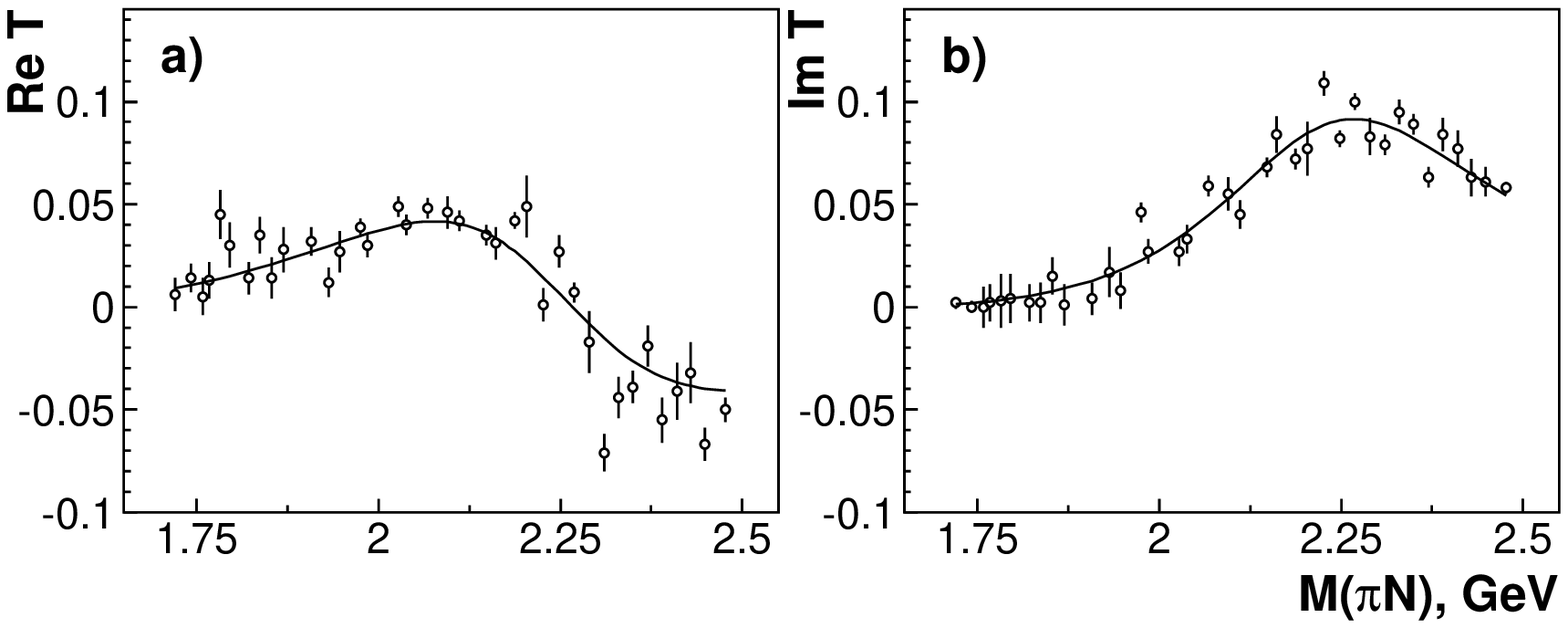,width=0.47\textwidth}
\epsfig{file=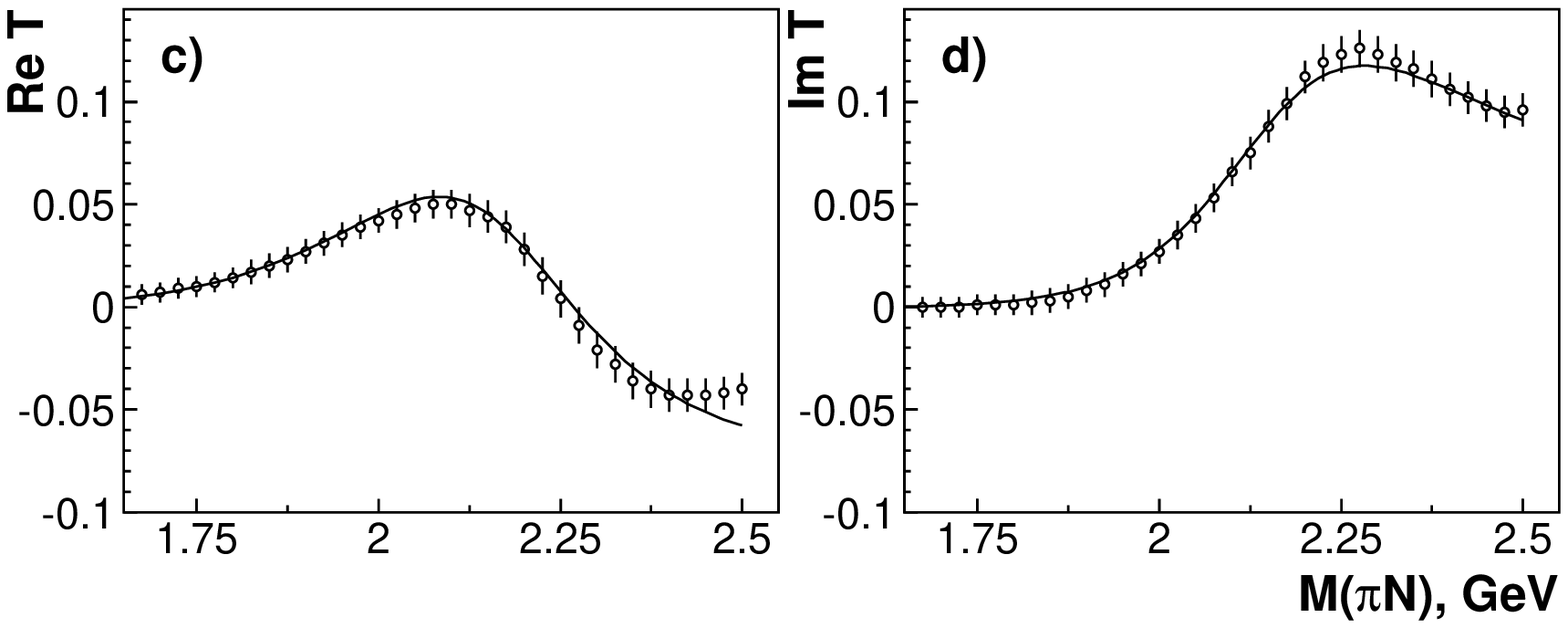,width=0.47\textwidth}
\caption{\label{pwa_g19}Real and imaginary part of the
$\frac12(\frac92) ^-$ partial wave.  Points with error bars are from
the SAID energy independent solution \cite{Arndt:2006bf} (a,b) and
from KH84 \cite{Hohler:1984ux} (c,d). For KH84, $\pm 5$\% errors
were assumed.}
\end{figure}

The state $N(2250)G_{19}$ has four stars by PDG classification and
is clearly seen in the elastic data. As in the case of the
$I(J^P)=1/2(7/2^+)$ wave, the data were fitted as one-pole
three-channel ($\pi N$, $\pi\Delta$ and $\rho N$) K-matrix with a
$\pi N\to \pi N$ non-resonant transition taken as a constant. The
description of the SAID and KA84 results is shown in
Fig.~\ref{pwa_g19}. The $N(2250)G_{19}$ state practically does not
contribute to the photoproduction data and no additional sensitivity
for this state was found when photoproduction data were included.
The $K\Lambda$ coupling also optimized at a rather small value and
this state contributed very little to the $\pi^- N\to K\Lambda$
data. Mass and width are given in Table~\ref{residuesneg}.

\section{Discussion and conclusions}
\subsection{Summary of nucleon resonances studied}

Four resonances which we observe in our analysis are new in the
sense that they have no defined entry in RPP
\cite{Nakamura:2010zzi}. We call them here $N_{1/2^+}(1880)$,
$N_{1/2^-}(1895)$, $N_{3/2^-}(1875)$, and $N_{5/2^-}(2060)$. All
four of them leave very significant traces in photoproduction, in
particular in the reaction $\gamma p\to K^+\Lambda$. Here, we would
like to remind the reader that nearly complete information exists on
this reaction: The differential cross sections are known with high
precision, the $\Lambda$ recoil polarization is determined from its
$N\pi$ weak decay, the beam asymmetry has been measured. Using
linearly and circularly polarized photons, the polarization transfer
coefficients from the initial photon to the $\Lambda$ hyperon in the
final state has been determined. From these quantities, the target
asymmetry can be deduced algebraically. With one further
measurement, e.g. the polarization correlation between target and
$\Lambda$ recoil polarization, the experimental information would be
complete, in the sense that the results of all other polarization
experiments can be predicted. Possibly, even a truly
energy-independent reconstruction of the partial wave amplitudes
will become possible.

\begin{figure}[pt]
\centerline{\epsfig{file=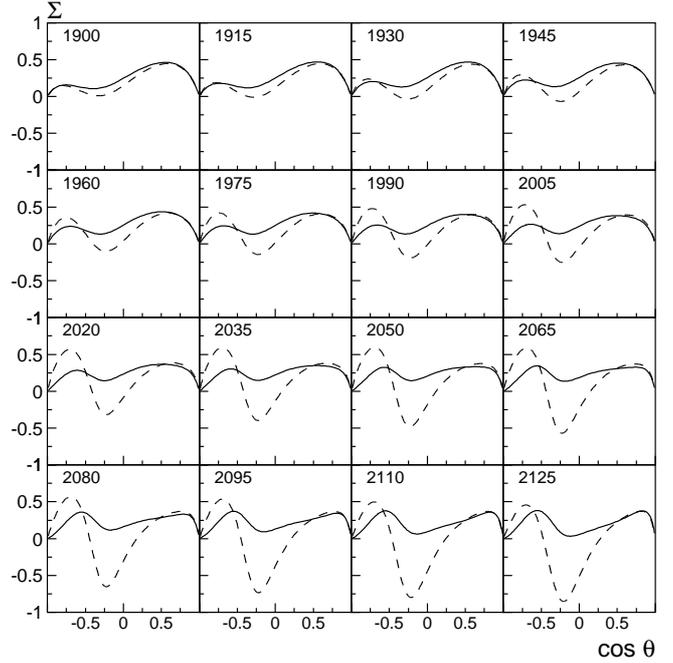,width=0.49\textwidth}}
\caption{\label{klam_pred1} Predictions of the beam asymmetry for
the $\gamma p\to K^+\Lambda$ reaction. The full curve corresponds to
the solution BG2011-02 and the dashed one corresponds to the
solution BG2011-01.}
\end{figure}
\begin{figure}[pt]
\centerline{\epsfig{file=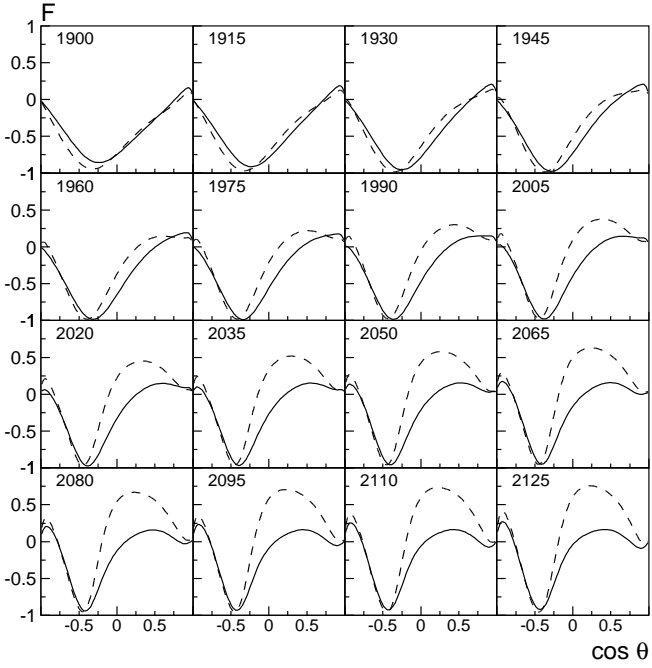,width=0.49\textwidth}}
\caption{\label{klam_pred2} Predictions of the double polarization
variable $F$ for the $\gamma p\to K^+\Lambda$ reaction. The full
curve corresponds to the solution BG2011-02 and the dashed one
corresponds to the solution BG2011-01.}
\end{figure}

In view of the excellent data base and the constraints provided by
an energy dependent fit, we believe the solution(s) we present are
highly constrained. To our judgement, the existence of these four
resonances ranges from very likely to certain even though further
information is desirable. At the same level we see the RPP
resonances $N(1900)P_{13}$ and $N(2080)D_{13}$  - the latter
resonance is observed at $M=2150\!\pm\!60$, $\Gamma
=330\!\pm\!45$\,MeV here. Based on pion-induced reactions, RPP
classified them as two-star resonances \cite{Nakamura:2010zzi}.

The evidence for the existence of the two resonances in $F$-wave,
the RPP states $N(2000)F_{15}$ and $N(1990)F_{17}$, is still only
fair. We hope that forthcoming data with higher sensitivity will
provide additional support for their existence and will define their
properties with improved accuracy. In Fig. \ref{klam_pred1} and
\ref{klam_pred2} we show the predicted beam asymmetry $\Sigma$  and
the double polarization variable $F$ - the correlation between the
degree of circular photon polarization and target polarization in
the scattering plane and perpendicular to the photon beam -  for
$\gamma p\to K^+\Lambda$ for our two solutions BG2011-01 and
BG2011-02. The figure may serve as an example that even a modest
accuracy in polarization observables can make a significant impact
on the results of a partial wave analysis.

\subsection{Interpretation}
For a discussion of the mass pattern of nucleon resonances, it is a
bit tedious to refer to a resonance as $N(2200)D_{15}$ and to state:
We, however, believe the mass of this resonance is rather 2075\,MeV
and not 2200\,MeV. Hence we list in Table \ref{multiplets}, with the
mass {\it we} find, all positive-parity resonances in the
1650-2250\,MeV mass range, and all negative-parity resonances in the
1750-2350\,MeV mass range.

\begin{table}[pt]
\caption{\label{multiplets}Light-quark nucleon resonances. For
resonances reported in this article we give our Breit-Wigner masses,
for well known states we give nominal RPP masses, and for states
reported in \cite{Anisovich:2010an} we use names given there
($M_{\rm eff}$). Mass values obtained from alternative solutions are
given in a separate line.}
\begin{scriptsize}
\renewcommand{\arraystretch}{1.20}
\begin{tabular}{ccccc}
\hline\hline\\[-2.ex]
\hspace{-5mm}$J =\quad 1/2$\qquad&\hspace{-3mm}$3/2$         &\hspace{-3mm}     $5/2$       &\hspace{-3mm}    $7/2$        &$9/2$\\[0.5ex] \hline\\[-2.ex]
\hspace{-1mm}$N_{1/2^+}(1710)$&                 &                 &                 &\\
                 &\hspace{-3mm}$N_{3/2^+}(1720)$&\hspace{-3mm}$N_{5/2^+}(1680)$&                 &\\
\hspace{-1mm}$N_{1/2^+}(1880)$&\hspace{-3mm}$N_{3/2^+}(1910)$&\hspace{-3mm}$N_{5/2^+}(2095)$&\hspace{-3mm}$N_{7/2^+}(2100)$&\\[-0.5ex]
     &    \tiny\qquad  alternatively:                              &\hspace{-3mm}\tiny$\sim\hspace{-1mm}1875+2200$&\hspace{5mm}\tiny$1990$&\\
                 &                 &                 &                 &\hspace{-3mm}$N_{9/2^+}(2220)$\\[0.5ex] \hline\\[-2.ex]
\hspace{-1mm}$N_{1/2^-}(1895)$&\hspace{-3mm}$N_{3/2^-}(1875)$&                 &                 &\\
                 &\hspace{-3mm}$N_{3/2^-}(2150)$&\hspace{-3mm}$N_{5/2^-}(2075)$&\hspace{-3mm}$N_{7/2^-}(2185)$&\hspace{-3mm}$N_{9/2^-}(2280)$\\[1.ex]
\hline\hline\\
\end{tabular}
\renewcommand{\arraystretch}{1.0}
\end{scriptsize}\vspace{-5mm}
\end{table}

Two comments have to be made:

i) For the interpretation we use Breit-Wigner masses instead of pole positions. The reason
is phenomenology. In lattice calculations and in quark models
neglecting decays, the masses of isospin doublets like ($\rho$,
$\omega$) and ($a_1(1260)$, $b_1(1235)$) are about mass degenerate.
The pole masses are different - in particular for the $J^P = 1^+$
doublet with the very wide $a_1(1260)$ - while the Breit-Wigner
masses are similar: The Breit-Wigner masses are closer to the bare
poles of resonances; to the mass of resonances before they ``dress
themselves with a meson cloud".

ii) The interpretations given below depend on the PWA solution. In a
given interpretation, we chose the most appropriate pattern of
resonances from Table \ref{multiplets}. These alternatives can be
decided as soon as new data can differentiate between the solutions
presented above.

\subsubsection{Interpretation within the quark model}
Within the quark model light-quark baryons are assigned to SU(6)
multiplets. For this discussion, some well known SU(6) relations are
needed which - for the convenience of the reader - are reproduced
here. The symmetry group SU(3) of the three flavors and the spin
symmetry group SU(2) combine to the group SU(6) which decomposes
according to
\be\label{6}
6\otimes6\otimes6 = 70_M + 70_M + 56_S +20_A
\ee
where the subscript denotes the symmetry behavior with respect to
the exchange of two quarks in the baryon. The baryon wave function
can be symmetric ($S$), antisymmetric ($A$), or of mixed symmetry
($M$). These SU(6) groups can be written as sum of SU(3) groups with
a defined spin multiplicity (given as superscript):
\begin{eqnarray}\label{70} 70 &=& ^210 + ^48 + ^28 + ^28 + ^21\\
56&=& ^410+^28\label{56} \\ 20&=&^28+^28+^41\label{1}
\end{eqnarray}

The two resonances $N_{3/2^+}(1720)$ and $N_{5/2^+}(1680)$ are
easily identified as spin doublet with intrinsic orbital and spin
angular momenta $L=2, S=1/2$. Spin and flavor wave functions both
have mixed symmetry, the spin-flavor \,wave function and the spatial
wave function are likely symmetric. Hence they can be assigned to
the (D,$L^P_{\,\rm N})=(56,2^+_2)$ multiplet. Here, D is the
dimensionality of the SU(6) multiplet, $L^P$ intrinsic orbital
angular momentum and the parity, N the shell number in the harmonic
oscillator classification. The four states
\be\label{quartet-p}
N_{1/2^+}(1880), N_{3/2^+}(1910), N_{5/2^+}(1875), N_{7/2^+}(1990)
\ee
form a spin quartet which we assign to the (D,$L^P_{\,\rm
N})=(70,2^+_2)$ multiplet. Here, we have chosen the alternative mass
values in Table \ref{multiplets} which are better adapted to the
quark model. Missing is a spin doublet (D,$L^P_{\,\rm
N})=(70,2^+_2)$ with $J^P=\frac32^+, \frac52^+$.

Alternatively, we may assume that we have a spin-quartet of nucleon
resonances
\be\label{quartet-h}
N_{1/2^+}(2100), N_{3/2^+}(1970), N_{5/2^+}(2095), N_{7/2^+}(2100)
\ee
and a spin-doublet
\be\label{doublet-p}
N_{1/2^+}(1880), N_{3/2^+}(1910). \ee

The doublet should have $L=1, S=1/2$ and positive parity. Such a
doublet is expected in quark models and assigned to a 20-plet in the
SU(6)$\times$O(3) classification. This expectation is confirmed in
recent lattice calculations \cite{Edwards:2011jj}. The two intrinsic
oscillators both carry angular momentum $l_i=1$, with $\vec L=\vec
l_1+\vec l_2$: excitation of both oscillators is apparent.

Two  positive-parity states, $N_{1/2^+}(1440)$ and
$N_{9/2^+}(2220)$, remain to be discussed:

The Roper resonance $N_{1/2^+}(1440)$ has been debated often; here
we interpret it as nucleon radial excitation in the (D,$L^P_{\,\rm
N})=(56,0^+_2)$ multiplet. This interpretation does not exclude a
sizable molecular component in the wave function, in particular at
small momentum transfer. The $N_{1/2^+}(1710)$ resonance could
belong to the (D,$L^P_{\,\rm N})=(70,0^+_2)$ multiplet. In this
interpretation we should expect an additional resonance with
$J^P=\frac32^+$. Finally, a further spin doublet is expected in
quark models with $J^P=\frac12^+, \frac32^+$, where two internal
quark angular momenta $l_{i}; i=1,2$ add to $L=1$ and which would
belong to the (D,$L^P_{\,\rm N})=(20,1^+_2)$ multiplet.

$N_{9/2^+}(2220)$ must have a minimum orbital angular momentum
$L=4$; its mass indicates where resonances in the fourth shell
should be expected. It is hence assigned to the (D,$L^P_{\,\rm
N})=(56,4^+_4)$ multiplet.

The negative-parity nucleon resonances of Table \ref{multiplets}
seem to be organized into a spin doublet
\be\label{doublet}
N_{1/2^-}(1895), N_{3/2^-}(1875) \ee with $L=1$ and a spin quartet
\be\label{quartet-n}
N_{3/2^-}(2150), N_{5/2^-}(2075), N_{7/2^-}(2190), N_{9/2^-}(2250)
\ee
with $L=3$. A spin doublet with $L=1$ in a 70-plet should be
accompanied by a spin triplet, i.e. at least a nucleon resonance
with $J^P=\frac52^-$ should exist at a close-by mass.
$N_{5/2^-}(2075)$ seems be too high in mass and rather to belong to
the spin-quartet. (The two additional resonances with
$J^P=\frac12^-$ and $J^P=\frac32^-$ expected in addition to form a
complete triplet could be hidden by a stronger production of the
$N_{1/2^-}(1895)$, $N_{3/2^-}$ $(1875)$ spin doublet.) Hence we
believe that $N_{1/2^-}(1895)$, $N_{3/2^-}$ $(1875)$ forms a spin
doublet. As spin doublet without accompanying triplet it must belong
to the (D,$L^P_{\,\rm N})=(56,1^-_3)$ multiplet. Then the quartet
(\ref{quartet-n}) must belong to a (D,$L^P_{\,\rm N})=(70,3^-_3)$
multiplet with a doublet with $J^P=\frac52^-$ and $\frac72^-$
missing. Possibly, they are unresolved parts of $N_{5/2^-}(2075)$,
$N_{7/2^-}$ $(2190)$. With these assignments, six SU(6) multiplets
are missing.

\subsubsection{Interpretation within quark-diquark model}

In the classical quark model, the three quarks obey the Pauli
principle. In conventional diquark models it is assumed that two
quarks -- forming the diquark -- are in  relative S wave. Under
these assumptions, the quartet of the states (\ref{quartet-p}) in
the mass region below 2 GeV is forbidden. However, when the wave
function of the two quarks in the diquark and the single quark have
no overlap, the Pauli principle should not applied: far separated
fermions do not obey the Pauli principle.

In \cite{Anisovich:2010wx} it is suggested that for highly excited
states, the quark-diquark wave function and the quark wave function
are spatially well separated. It is then an experimental question,
which orbital angular momentum is required to satisfy the criterium
of ``full separation". If for $L=2$ the separation is already
sufficiently large, a quartet of states (\ref{quartet-p}) is
predicted in the region 1.9 GeV. At low energies, where the system
is more compact and the quark-diquark wave functions overlap
strongly, the baryon wave function obeys fully the Pauli principle
and the additional states are suppressed.

\subsubsection{Interpretation within AdS/QCD}

AdS/QCD is an analytically solvable ``gravitational" theory
simulating QCD which is defined in a five-dimensional Anti-de Sitter
(AdS) space embedded in six dimensions
\cite{deTeramond:2005su,Karch:2006pv,Forkel:2007cm,Brodsky:2010ev}.
In a special variant of AdS/QCD, a two-parameter mass formula was
derived \cite{Forkel:2008un} which reproduces the baryon mass
spectrum with a surprising precision. For the six rows in Table
\ref{multiplets}, the predicted masses are 1735, 1735, 1926, 2265,
1833, 2102\,MeV, respectively. We mention that the mass formula had
been suggested before on an empirical basis \cite{Klempt:2002vp}.
Again, the alternative solutions in Table \ref{multiplets} are
preferred.

\subsubsection{Interpretation as parity doublets}

Table \ref{multiplets} exhibits a striking number of parity
doublets. These are collected in Table \ref{pd}. To the nucleons of
the first line in Table \ref{multiplets} we have added the triplet
of negative-parity resonances belonging to the first excitation
band. The mass values of ambiguous solutions are chosen to match the
masses of their negative-parity partners.
\begin{table}[pb]
\caption{\label{pd}Nucleon resonances as parity doublets. The masses
of states for which ambiguous solutions exist are chosen to match
their negative parity partners. Definition of masses as in
Table~\ref{multiplets}.}
\renewcommand{\arraystretch}{1.4}
\begin{tabular}{ccccc}\hline\hline
$N_{1/2^+}(1710)$&\hspace{-2mm}$N_{1/2^-}(1650)$&&\hspace{-1mm}$N_{3/2^+}(1720)$&\hspace{-2mm}$N_{3/2^-}(1700)$\\
$N_{5/2^+}(1680)$&\hspace{-2mm}$N_{5/2^-}(1675)$&&\hspace{-1mm}$N_{1/2^+}(1880)$&\hspace{-2mm}$N_{1/2^-}(1895)$\\
$N_{3/2^+}(1910)$&\hspace{-2mm}$N_{3/2^-}(1875)$&&\hspace{-1mm}$N_{5/2^+}(2095)$&\hspace{-2mm}$N_{5/2^-}(2075)$\\
$N_{7/2^+}(2100)$&\hspace{-2mm}$N_{7/2^-}(2190)$&&\hspace{-1mm}$N_{9/2^+}(2220)$&\hspace{-2mm}$N_{9/2^-}(2250)$\\\hline\hline
\end{tabular}
\end{table}

The occurrence of parity doublets in the mass spectrum of mesons and
baryons is surprising. The harmonic oscillator states with positive
and negative parity alternate, and in quark models, even in fully
relativistic quark models, this pattern survives. In meson
spectroscopy, a large number of resonances comes in nearly mass
degenerate parity doublets, however with important exemptions:
Mesons like $f_2(1270)$ and $a_2(1320)$ with ${\mathtt
J^{\mathtt{PC}}=2^{\mathtt{++}}}$, $\omega_3(1670)$ and
$\rho_3(1690)$ with ${\mathtt J^{\mathtt{PC}}}=$ $3^{\mathtt{--}}$,
$f_4(2050)$ and $a_4(2040)$ with ${\mathtt
J^{\mathtt{PC}}=4^{\mathtt{++}}}$, none of these states falling onto
the leading Regge trajectory has a mass-degenerate spin-parity
partner. These are mesons in which the orbital angular momentum
${\mathtt L}$ and the total quark spin ${\mathtt S}$ are aligned to
give the maximal ${\mathtt J}$ and which have the lowest mass in
that partial wave. Their chiral partners have considerably higher
masses: $\eta_2(1645)$ and $\pi_2(1670)$ (${\mathtt
J^{\mathtt{PC}}=2^{\mathtt{-+}}}$), $h_3(2045)$ and $b_3(2035)$
(${\mathtt J^{\mathtt{PC}}=3^{\mathtt{+-}}}$), $\eta_4(2320)$ and
$\pi_4(2250)$ (${\mathtt J^{\mathtt{PC}}}=$ $4^{\mathtt{-+}}$),
respectively \cite{Bugg:2004xu}. A graphical illustration is given
in Fig.~1 of \cite{Afonin:2006wt} and Fig.~57 of
\cite{Klempt:2007cp}. In \cite{Glozman:2010rp} it is argued that
formation of the spin-parity partners of mesons on the leading Regge
trajectory could be suppressed by angular momentum barrier factors.
A reanalysis of the reaction $\bar pp\to \pi\eta\eta$ in flight
\cite{Anisovich:2000ut} was performed and a weak indication claimed
\cite{Glozman:2010rp} for the possible existence of the missing
$4^{-+}$ state $\eta_4(1950)$ at about 1.95\,GeV. However, the
weakness of the signal is certainly not enforcing any
interpretation. Nevertheless, the suppression in $\bar pp$ formation
of spin-parity partners of mesons on the leading Regge trajectory is
certainly an argument which reduces the weight of their
non-observation.

In baryon spectroscopy, a similar phenomenon occurs. Even-parity
resonances with $S=3/2$ on a leading Regge trajectory seem to have
no spin-parity partner \cite{Klempt:2002tt}. In formation
experiments, the same mechanism as in meson formation could possibly
suppress these states. In elastic scattering, an angular momentum
${\mathtt L}=3$ is required to form a $7/2^+$ resonance; for
$7/2^-$, ${\mathtt L}=4$ is needed. In photoproduction of baryon
resonances, on the contrary, such a suppression is not expected.
Baryons on the leading Regge trajectory, e.g. a $7/2^+$ resonance
with the lowest mass, requires a $E_4 ^+$ or $M_4 ^+$ amplitude, a
$7/2^-$ resonance a $E_3^-$ or $M_3^-$ amplitude. There is no
kinematical factor which would suppress production of $7/2^-$
resonances compared to $7/2^+$ resonances. Photoproduction
experiments could thus be of decisive importance to clarify the
dynamics of highly excited hadrons. Unfortunately, the two solutions
for the mass of the $5/2^+$ and $7/2^+$ resonances (listed in Table
\ref{multiplets} prevent that a final decision can be made at
present. Future experiments and analyses will be required. But the
predictions in Fig. \ref{klam_pred1} and \ref{klam_pred2} show that
a decision might be close.

\subsection{Summary}
In a multichannel partial wave analysis of nearly all existing data
on pion- and photo-induced reactions we find evidence for a number
of new or badly established resonances, and determine their
properties. In particular, we observe four new resonances
$N_{1/2^+}(1875)$, $N_{1/2^-}(1895)$, $N_{3/2^-}(1875)$ and
$N_{5/2^-}$ $(2075)$, and two resonances, $N(1900)P_{13}$ and
$N(2080)D_{13}$ which were badly known so far. We believe that the
existence of these three resonances is almost certain even though
further confirmation is desirable. The new (and old) resonances are
interpreted within different models.

\subsection*{Acknowledgements}
We would like to thank the members of SFB/TR16 for continuous
encouragement. We acknowledge support from the Deutsche
Forschungsgemeinschaft (DFG) within the SFB/ TR16 and from the
Forschungszentrum J\"ulich within the FFE program.

\end{document}